\documentclass[10pt,journal,compsoc]{IEEEtran}

\ifCLASSOPTIONcompsoc

\usepackage[nocompress]{cite}
\else
\usepackage{cite}
\fi
\usepackage{longtable}
\usepackage{booktabs}
\usepackage{amsmath}
\usepackage{graphicx}
\usepackage{mwe}
\usepackage{graphbox}
\usepackage{amsfonts} 
\usepackage{enumitem}
\usepackage{amssymb}
\usepackage{ragged2e}
\usepackage[colorlinks, linkcolor=blue, anchorcolor=blue, citecolor=blue]{hyperref}
\usepackage[numbers,sort&compress]{natbib}
\usepackage{comment}
\usepackage{color}
\usepackage{bbding} 
\usepackage{diagbox} 
\usepackage{stfloats}
\usepackage{float}
\usepackage{multirow}
\usepackage{subfigure}
\usepackage[justification=centering]{caption}
\usepackage{ntheorem}

\usepackage{subcaption}

\usepackage[export]{adjustbox}
\theorembodyfont{\upshape}

\newtheorem{theorem}{Theorem}[section] 
\newtheorem{example}{Example}[section]
 
\newtheorem*{proof}{Proof}
\newtheorem{definition}[theorem]{Definition} 

\theoremstyle{remark}
\ifCLASSINFOpdf

\ifCLASSINFOpdf

\else

\fi

\begin{document}

\title{QKSAN: A Quantum Kernel Self-Attention Network}

\author{Ren-Xin Zhao,~\IEEEmembership{Member,~IEEE,} Jinjing Shi\textsuperscript{*},~\IEEEmembership{Senior Member,~IEEE,} and~Xuelong~Li,~\IEEEmembership{Fellow,~IEEE}
\IEEEcompsocitemizethanks{\IEEEcompsocthanksitem Ren-Xin Zhao is with the School of Computer Science and Engineering, Central South Univerisity, China, Changsha, 410083. Jinjing Shi is with the School of Electronic Information, Central South Univerisity, China, Changsha, 410083. 
\IEEEcompsocthanksitem Xuelong Li is with the School of Artificial Intelligence, OPtics and ElectroNics, Northwestern Polytechnical University, China, Xi’an, 710072.
\IEEEcompsocthanksitem{Jinjing Shi is the corresponding author.} 
\IEEEcompsocthanksitem{E-mails: shijinjing@csu.edu.cn, renxin\_zhao@alu.hdu.edu.cn, li@nwpu.e
	du.cn.} 
}
\thanks{Manuscript received XXXX; revised XXXX.}}


\IEEEtitleabstractindextext{%
\begin{abstract}\justifying
Self-Attention Mechanism (SAM) excels at distilling important information from the interior of data to improve the computational efficiency of models. Nevertheless, many Quantum Machine Learning (QML) models lack the ability to distinguish the intrinsic connections of information like SAM, which limits their effectiveness on massive high-dimensional quantum data. To tackle the above issue, a Quantum Kernel Self-Attention Mechanism (QKSAM) is introduced to combine the data representation merit of Quantum Kernel Methods (QKM) with the efficient information extraction capability of SAM. Further, a Quantum Kernel Self-Attention Network (QKSAN) framework is proposed based on QKSAM, which ingeniously incorporates the Deferred Measurement Principle (DMP) and conditional measurement techniques to release half of quantum resources by mid-circuit measurement, thereby bolstering both feasibility and adaptability. Simultaneously, the Quantum Kernel Self-Attention Score (QKSAS)  with an exponentially large characterization space is spawned to accommodate more information and determine the measurement conditions. Eventually, four QKSAN sub-models are deployed on PennyLane and IBM Qiskit platforms to perform binary classification on MNIST and Fashion MNIST, where the QKSAS tests and correlation assessments between noise immunity and learning ability are executed on the best-performing sub-model. The paramount  experimental finding is that a potential learning advantage is revealed in partial QKSAN subclasses that acquire an impressive more than 98.05\%  high accuracy with very few parameters that are much less in aggregate than classical machine learning models.  Predictably, QKSAN lays the foundation for future quantum computers to perform machine learning on massive amounts of data while driving advances in areas such as quantum computer vision. 
\end{abstract}


}

\maketitle

\IEEEdisplaynontitleabstractindextext

\IEEEpeerreviewmaketitle

\section{Introduction}\label{introduction}

\IEEEPARstart{I}{n} recent years, QML has developed tremendously \cite{0.02,0.021,0.022}. However, many current QML models treat each quantum data equally and neglect the value of the intrinsic connections between data, which demands large amounts of expensive quantum storage to remember all the information while hampering future large-scale quantum data processing on quantum computers. If in the classical domain, the above urgent issues can be effectively addressed by SAM. 

SAM was originally proposed in 2017 \cite{0.0} as a module that generates new sequences by calculating self-attention scores from the sequences themselves, which allows machine learning models to better capture crucial messages in the sequences while relying less on external information, thereby dramatically boosting the processing efficiency of models. To date, SAM has accomplished impressive achievements in natural language processing \cite{0.01,0.03}, computer vision \cite{0.01,0.04,0.041,0.042}, recommender system \cite{0.05,0.06,0.07} and other fields with the above advantages. One of the prominent instances is that the application of global SAM has enabled the BoTNet model to realize 84.7\% top-1 accuracy in the ImageNet benchmark test \cite{0.1}. Albeit very powerful, the time and memory complexity of classical SAM \cite{0.0} scales quadratically with sequence length, thus hampering its exploitation in domains with long contexts. A new idea for dilemma of SAM is offered by kernel methods. In 2019, a SAM based on kernel methods was first introduced, paving the way for designing attention using a larger representation space \cite{0.2}. In 2021, SAM was substituted with a feature mapping that can directly approximate the softmax kernel, enabling the use of pre-trained weights in linear time models \cite{0.4}. The above two cases substantiate strongly that kernel methods not only outperform in data representation efficiency, but also surmount the high-complexity bottleneck, thereby ushering in novel directions for the SAM paradigm. The success of kernel methods in SAM naturally sparks interest in QKM.

QKM, a new variant of kernel methods, are receiving increasing attention as a bridge between quantum computing and classical machine learning theory. It can embed classical data into ansatzes that are difficult to simulate classically \cite{0.5,0.6}. This embedding process is equivalent to mapping the data nonlinearly into quantum Hilbert space, thus stimulating quantum advantages \cite{0.7,0.71}. Here, the quantum advantage refers to the performance of QKM beyond its classical counterpart, which is reflected in multiple aspects. Firstly, QKM are better at finding globally optimal solutions than quantum neural networks in some supervised machine learning tasks \cite{0.71}. Secondly, QKM exhibit exponential acceleration for specific problems that are classically arduous to generate correlations \cite{0.8}, such as encoding inductive bias \cite{0.10}, resulting in remarkable learning capabilities. Moreover, its exceptional universal approximation properties and expressiveness enable QKM-based QML models to handle a wide range of classical machine learning tasks while efficiently addressing bounded-error quantum polynomial-time problem \cite{0.101,0.11,0.9}. 

The above statements prompt a natural inquiry into whether the merits of QKM and SAM can be merged to yield new models to further strengthen the data representation and information extraction capabilities of the QML framework for undifferentiated training data. The subsequent pivotal challenge is how to present the combination of QKM and SAM mechanisms in the form of quantum circuits. The last concern is how effective the framework is in practice. To this end, the \textbf{contribution} of this paper lies in addressing these pertinent inquiries and presenting novel insights into the potential of this novel QML framework:

\begin{itemize}
	\item From a quantum perspective, the QKSAM theory is proposed to establish the QKM and SAM connection.
	
	\item Led by above theory, based on DMP and conditional measurement techniques, QKSAN framework for the binary classification task of MNIST and Fashion MNIST is constructed to strengthen the applicability and feasibility.
	
	\item Finally, Angle encoding Hardware-Efficient ansatz (AnHE), Amplitude encoding Hardware-Efficient ansatz (AmHE), Angle encoding QAOA ansatz (AnQAOA) and Amplitude encoding QAOA ansatz (AmQAOA) are deployed on  PennyLane and IBM Qiskit platforms to reveal the most valuable potentials that an impressive  more than 98.05\% high accuracy is attained by the partial QKSAN subclasses with very few parameters that are much less in amount than classical machine learning models. 
\end{itemize}

\begin{table*}[h]
	%
	\def\tablename{Tab.}
	\centering
	\caption{Quantum Gates}
	\label{Notations}
	\small
	\begin{tabular}{@{}lccc@{}}
		\toprule
		Name of Quantum Gate & Mathematical Notation & Matrix Representation & Symbol of Quantum Circuit \\ \midrule
		
		Pauli X gate& $R_{X}$& $\left[ \begin{matrix}
 \cos\left(\frac{\theta}{2}\right)   & -i\sin\left(\frac{\theta}{2}\right) \\ \rule{0pt}{12pt} 
-i\sin\left(\frac{\theta}{2}\right) & \cos\left(\frac{\theta}{2}\right)\\
		\end{matrix} \right]$
		
		&         
		\includegraphics[align=c,scale=0.126]{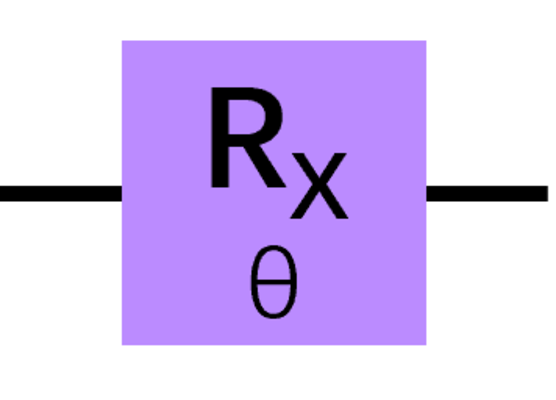} \\ \rule{0pt}{22pt}
		
		Pauli Y gate& $R_{Y}$& $\left[ \begin{matrix}
			\cos \left( \frac{\theta }{2} \right) & -\sin \left( \frac{\theta }{2} \right)  \\ \rule{0pt}{12pt} 
			\sin \left( \frac{\theta }{2} \right) & \cos \left( \frac{\theta }{2} \right)  \\
		\end{matrix} \right]$

		&         
		\includegraphics[align=c,scale=0.126]{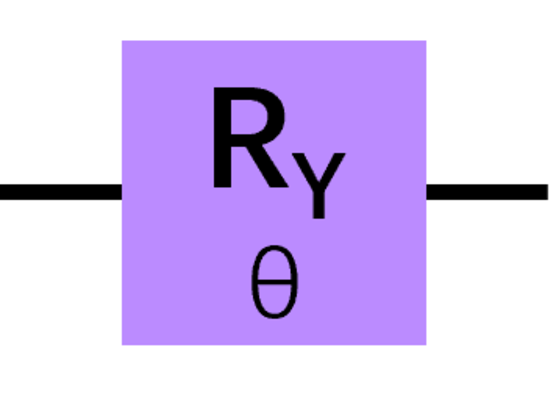}   \\ \rule{0pt}{22pt}
		
		Pauli Z gate & $R_{Z}$&   $\left[ \begin{matrix}
 e^{-i\frac{\theta}{2}} & 0 \\ \rule{0pt}{12pt} 
0 & e^{i\frac{\theta}{2}}\\ 
		\end{matrix} \right]$

		&    \includegraphics[align=c,scale=0.126]{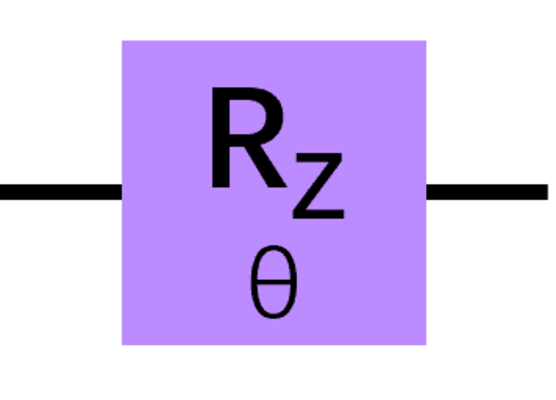}             \\  \rule{0pt}{32pt}         
		
		Controlled Y gate &   $CR_{Y}$& $\left[ \begin{matrix}
 1 & 0         & 0 & 0 \\\rule{0pt}{10pt} 
0 & \cos\left(\frac{\theta}{2}\right) & 0 & -\sin\left(\frac{\theta}{2}\right) \\\rule{0pt}{10pt} 
0 & 0         & 1 & 0 \\ \rule{0pt}{10pt} 
0 & \sin\left(\frac{\theta}{2}\right) & 0 & \cos\left(\frac{\theta}{2}\right) \\
		\end{matrix} \right]$

		&    \includegraphics[align=c,scale=0.132]{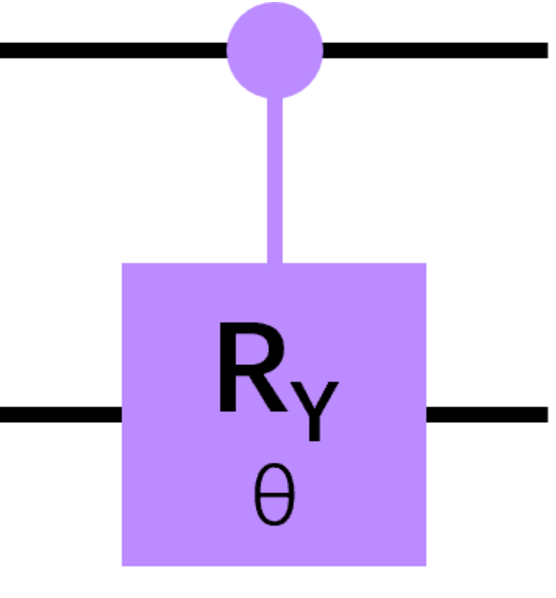}                    \\ \rule{0pt}{22pt}    
		
		Hadamard gate &    $H$ &$\frac{1}{\sqrt{2}}\left[ \begin{matrix}
			1 & 1  \\
			1 & -1  \\
		\end{matrix} \right]$
		&         \includegraphics[align=c,scale=0.1]{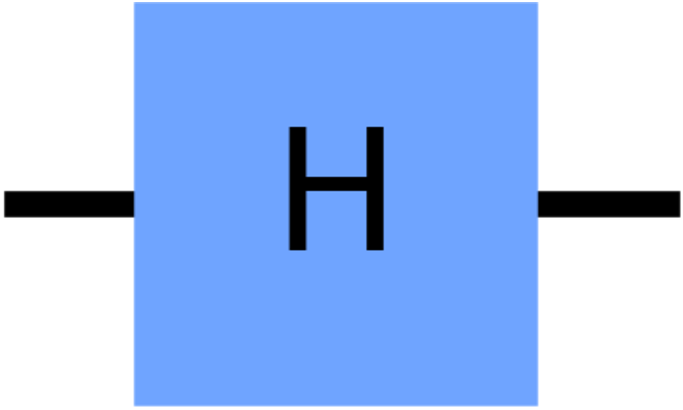}                                              \\ \rule{0pt}{28pt}

		CNOT gate&      $CNOT$ &  $\left[ \begin{matrix}
			1 & 0 & 0 & 0  \\
			0 & 1 & 0 & 0  \\
			0 & 0 & 0 & 1  \\
			0 & 0 & 1 & 0  \\
		\end{matrix} \right]$
		&     \includegraphics[align=c,scale=0.1]{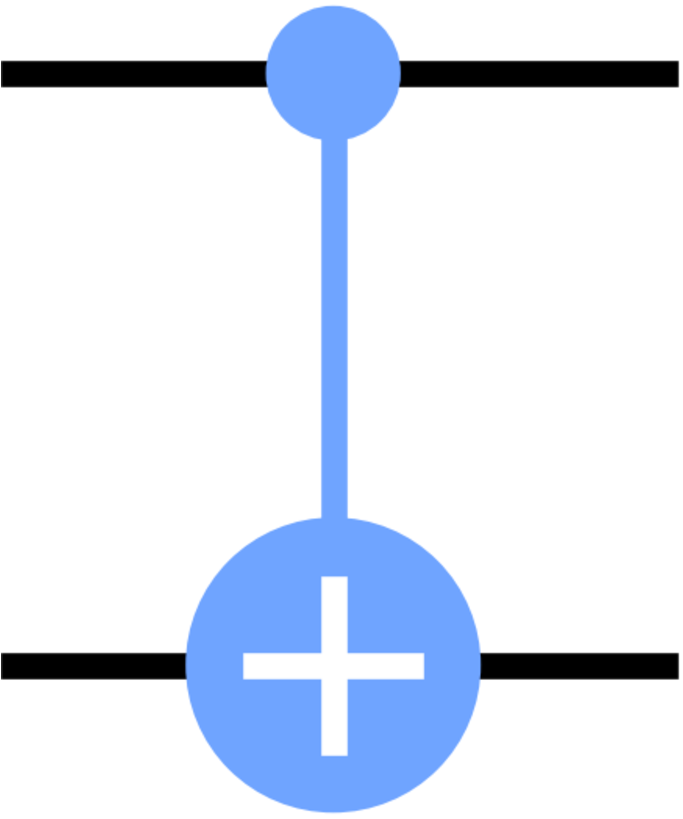}            \\

		\bottomrule
	\end{tabular}%
\end{table*}
The structure of the remainder of this paper is as follows: Quantum Attention Mechanisms (QAM), QKM and DMP are outlined in Section 2 to provide a preliminary understanding of QKSAN. In Section 3, QKSAM is portrayed mathematically. The QKSAN framework and its corresponding workflow are detailed in Section 4. Experiments are conducted and a series of meaningful  discoveries are obtained in Section 5. Eventually, in section 6, conclusions are drawn based on the findings presented in the preceding sections.

\section{Related Works}

In this part, a succinct exposition is presented concerning  QAM, the underpinnings of quantum computing, SAM, QKM, and DMP, serving as the bedrock for QKSAN.
\subsection{Quantum Attention Mechanisms}\label{QAM}

Since its introduction, SAM \cite{0.0} has exhibited a profound impact across various domains. Similarly, in QML, QAM seek to amplify QML model performance by discerning quantum data significance. Early investigations into QAM are inspired by quantum physics principles. For instance, the parameter-free QAM employing weak measurements was introduced in 2017 to enhance bidirectional LSTM sentence modeling, yielding superior outcomes compared to traditional attention mechanisms \cite{0.900}. However, such studies lack explicit quantum circuit schemes and rely solely on classical computer simulations. It was not until 2022 that QAM with variational architectures came to the fore. A probabilistic full-quantum self-attention network, implemented entirely on a quantum computer with exponentially scalable self-attention representation, was proposed by Zhao et al \cite{0.901}. In similar setups, it outperforms the hardware-efficient ansatz and QAOA ansatz in terms of learning capacity. In the same year, a quantum self-attention network leveraging a hardware-efficient ansatz for constructing a quantum SAM was devised by a Baidu team \cite{0.902}, achieving notable success in text categorization. Despite these advancements, QAM are in its early research phase, necessitating further theoretical and practical development. Constrained by quantum hardware limitations, they face challenges in broad applications. However, their substantial developmental potential suggests they may introduce innovative avenues for large-scale quantum data processing.

\subsection{Quantum Computing}\label{Quantum gates}

A figurative presentation of quantum theory can be achieved through the qubits and quantum gates. Therein, a qubit is the smallest unit that carries information. It is usually represented by the Dirac notation 
\begin{equation}\label{qubit}
	|\psi \rangle =\alpha |0\rangle +\beta |1\rangle ,
\end{equation}
where ground state $|0\rangle ={{[1,0]}^{\text{T}}}$, excited state $|1\rangle ={{[0,1]}^{\text{T}}}$. $\alpha$ and $\beta$ are amplitudes satisfying $|\alpha|^2+|\beta|^2=1$ \cite{1.01}. As shown in Eq. (\ref{qubit}), a qubit is in a superposition of the ground state and the excited state. This property endows it with exponential data representation abilities and a host of probabilistic features, which sets it apart significantly from classical computing. Moreover, a quantum gate can linearly transform one quantum state into another, in mathematical form as a $2^n\times2^n$ unitary matrix, where $n$ is the number of qubits. The quantum gates used in this paper include Hadamard gate, Pauli X gate, Pauli Y gate, Pauli Z gate, CNOT gate and Controlled Y gate as shown in Tab. \ref{Notations}.

\subsection{Self-Attention Mechanism}\label{Aa}

SAM is the investigative background of this paper, where its input $\mathbf{In}=\{{{\mathbf{w}}_{i}}\}_{i=0}^{n-1}$, ${{\mathbf{w}}_{i}}\in {{\mathbb{R}}^{1\times l}}$ and the output $\mathbf{Out}=\{\mathbf{new\_}{{\mathbf{w}}_{j}}\}_{j=0}^{n-1}$, $\mathbf{new}\_{{\mathbf{w}}_{j}}\in {{\mathbb{R}}^{1\times l}}$ are defined, where $n$ is the total number of outputs, $l$ is the dimension of the element. $ {{\mathbf{w}}_{i}} $ ($ \mathbf{new}\_{{\mathbf{w}}_{j}} $) indicates the $ i $-th ($ j $-th) element of the sequence $ \mathbf{In} $ ($ \mathbf{Out}$). For conciseness, it is specified that all subscripts in the text are used to indicate the location of the variable in the sequence, except for special instructions.

	Based on the above, SAM \cite{0.0} can be stated as
	\begin{equation}\label{self_attent}
		\mathbf{new}\_{{\mathbf{w}}_{i}}=\sum\limits_{j}{{{w}_{i,j}}{{\mathbf{V}}_{j}}}.
	\end{equation}In Eq. (\ref{self_attent}), $\mathbf{new}\_{{\mathbf{w}}_{i}}$ represents new output after the weighting operation. The weights
	\begin{equation}\label{weight}
		{{w}_{i,j}}=\text{softmax}\left( \frac{{{\mathbf{Q}}_{i}}\mathbf{K}_{j}^{\text{T}}}{\sqrt{d}} \right),
	\end{equation}
	also called attention scores, are obtained by normalizing the inner product ${{\mathbf{Q}}_{i}}\mathbf{K}_{j}^{\operatorname{T}}$. $\sqrt{d}$ is a scaling factor. In Eq. (\ref{self_attent}) and Eq. (\ref{weight}),
	\begin{equation}\label{q1}
		{{\mathbf{Q}}_{i}}={{\mathbf{w}}_{i}}\cdot {{U}_{Q}},
	\end{equation} 
	\begin{equation}\label{v1}
		{{\mathbf{V}}_{j}}={{\mathbf{w}}_{j}}\cdot{{U}_{V}} ,
	\end{equation}
	and $\mathbf{K}_{j}^{\text{T}}$ are the query vector, the value vector and the transpose of the key vector
	\begin{equation}\label{k1}
		{{\mathbf{K}}_{j}}={{\mathbf{w}}_{j}}\cdot {{U}_{K}},
	\end{equation} where $\mathbf{w}_{i}$ and $\mathbf{w}_{j}$ are inputs. In Eq. (\ref{q1}) to Eq. (\ref{k1}), ${{U}_{Q}}$, ${{U}_{K}}$ and ${{U}_{V}}$ are three trainable parameter matrices named as query conversion matrix, key conversion matrix and value conversion matrix respectively. 


\subsection{Quantum Kernel Methods}\label{QKMS}

QKM, a fundamental mechanism in QKSAN, revolves around leveraging Quantum Feature Mapping (QFM) for resolving quantum state similarity via measurement. 
First, QFM is defined as follows.
	Let  $\cal{X}$ be a set of input data. A QFM $\phi$ is a function that acts as  $\phi : \cal{X} \rightarrow\cal{F}$, where  $\cal{F}$ is the Hilbert  space.  This map \cite{0.6} transforms  $x \rightarrow |\phi(x)\rangle$,  where $x \in \cal{X}$, by way of a unitary transformation  $U_{\phi}(x)$ which is typically an ansatz whose parameters depend on the input data, i.e.
	\begin{equation}\label{QKM0}
		|\phi(x)\rangle =U_{\phi}(x)|\mathbf{0}\rangle .
	\end{equation}
	\begin{figure}[ht]
	\centering\includegraphics[scale=0.23]{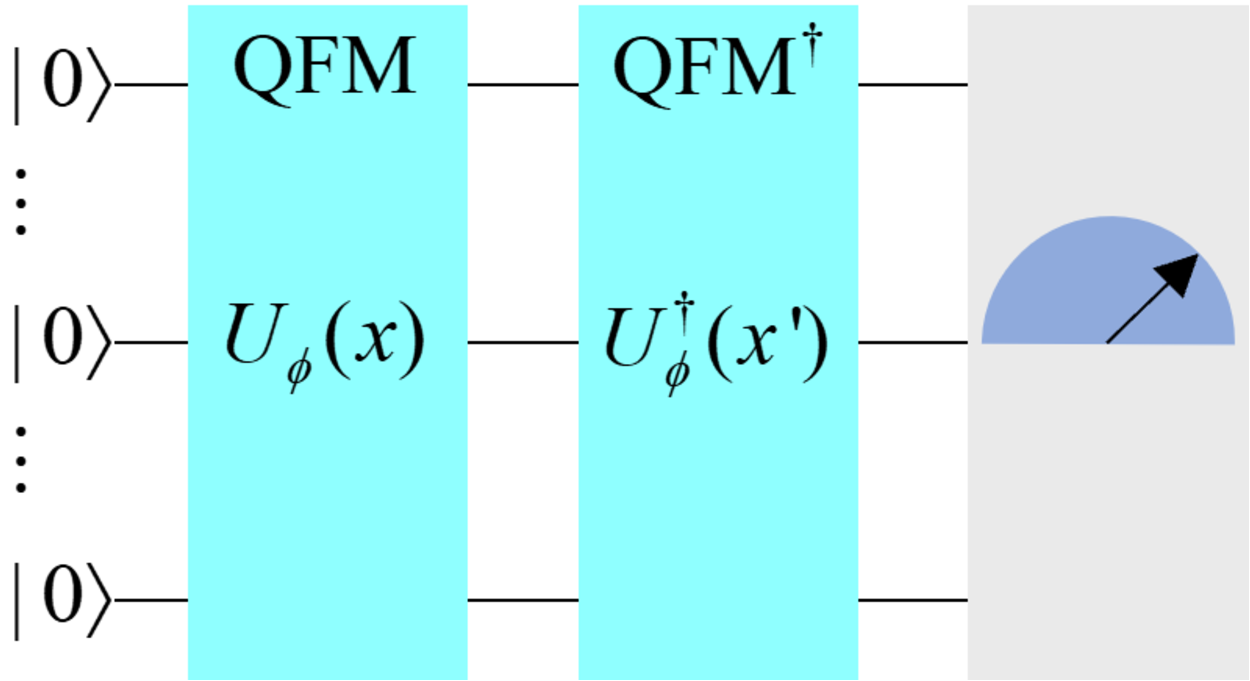}
	\caption{Quantum kernel estimation ansatz}\label{adjoint}
	\label{QKM2}
\end{figure}

	
Then, QKM evaluate the similarity 	
	\begin{equation}\label{QKM1}
		\kappa (x,x')=|\langle \phi(x) |\phi(x') \rangle {{|}^{2}}
	\end{equation}  using the quantum kernel estimation ansatz in Fig. \ref{QKM2} consisting of QFM circuit and its inverse circuit QFM$^{\dagger}$ embedded in the data point $x, x' \in \cal{X}$, respectively \cite{0.5}.

	\begin{figure*}[h]
		\centering
		\subfigure[Amplitude coding]{
			\centering\label{AMPLITUDU}\includegraphics[height=3.4cm]{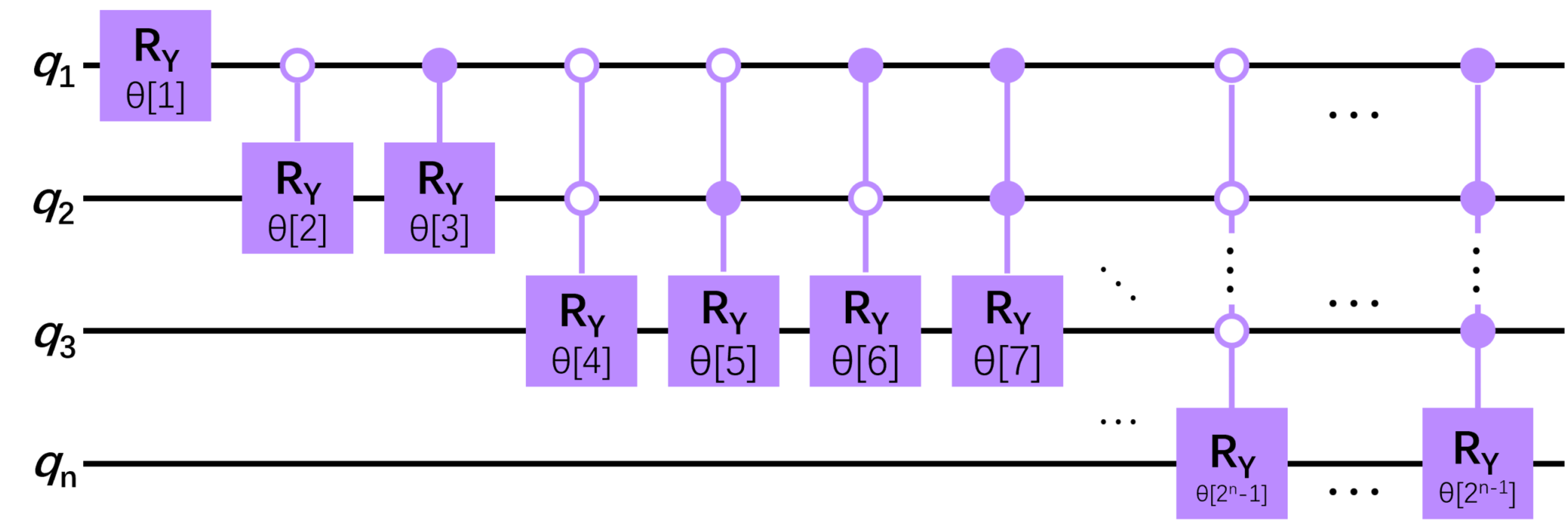}
		}
		\subfigure[Angle coding]{
			\centering\label{angles}\includegraphics[height=3.4cm]{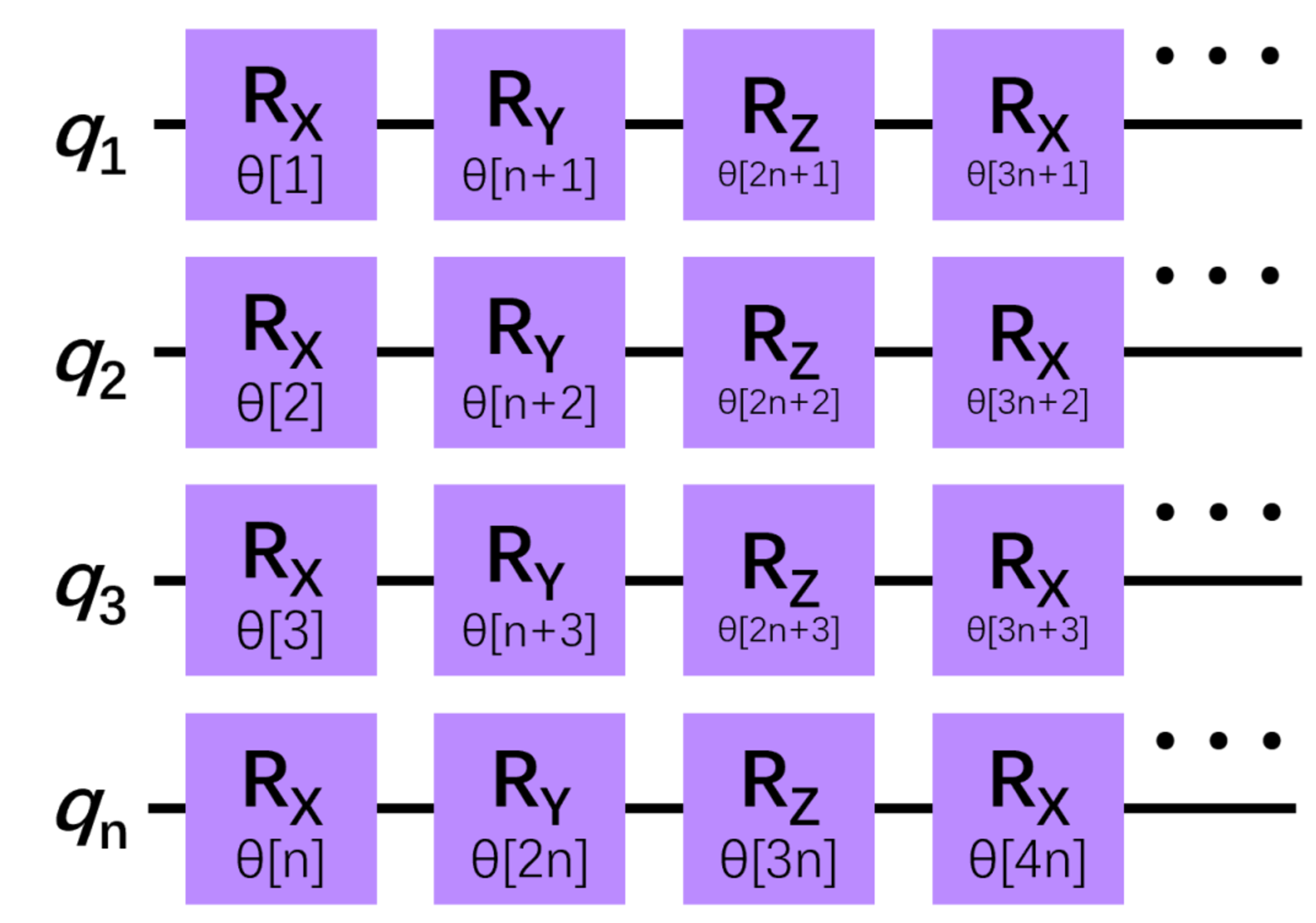}
		}
		\subfigure{
			\centering\label{valid1}\includegraphics[height=3.28cm]{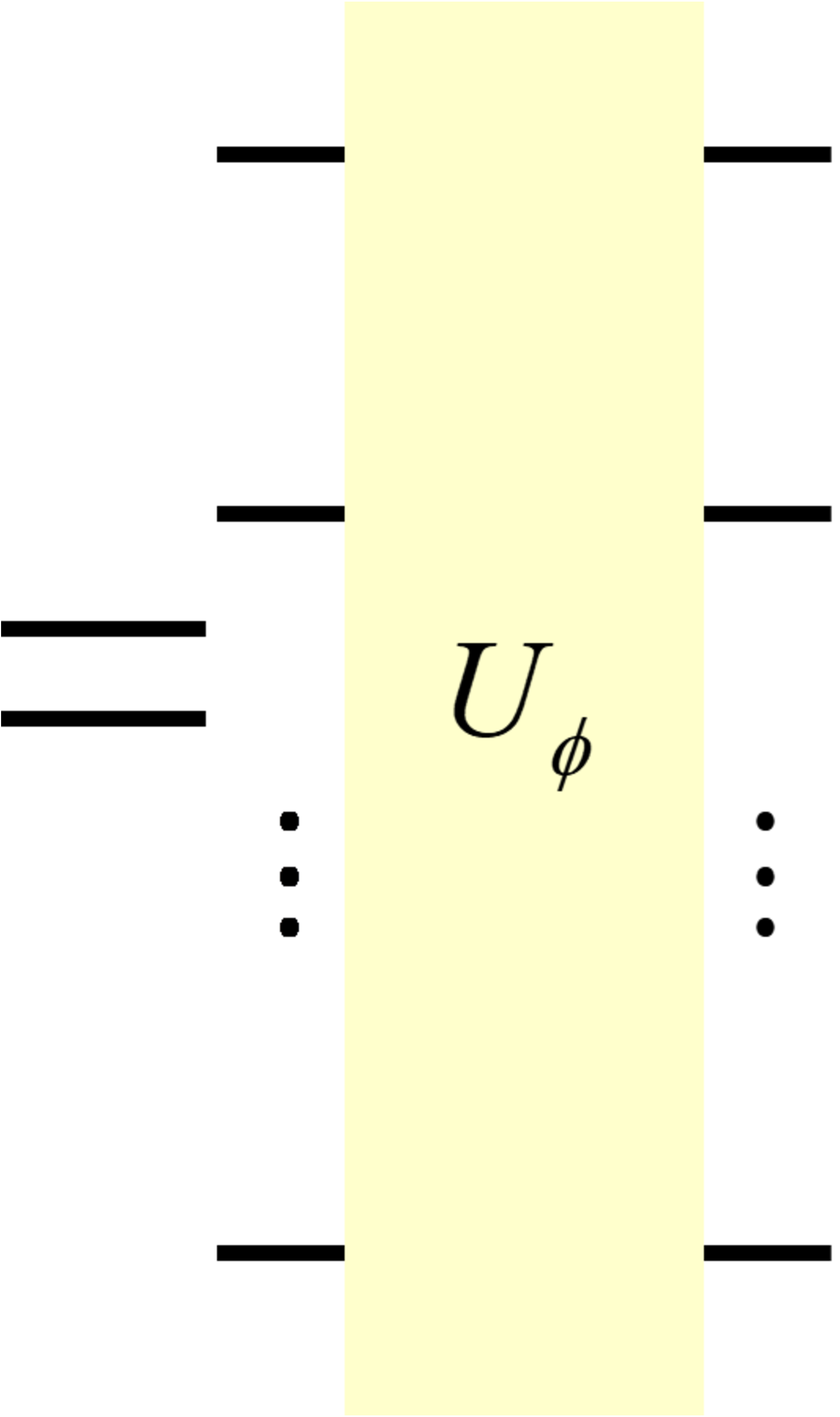}
	}
	\caption{The structure of ${{U}_{\phi }}$}\label{f0}
	
	\quad
	
	\centering
	\subfigure[QAOA ansatz]{
		\centering\label{QAOAansatz}\includegraphics[height=3.3cm]{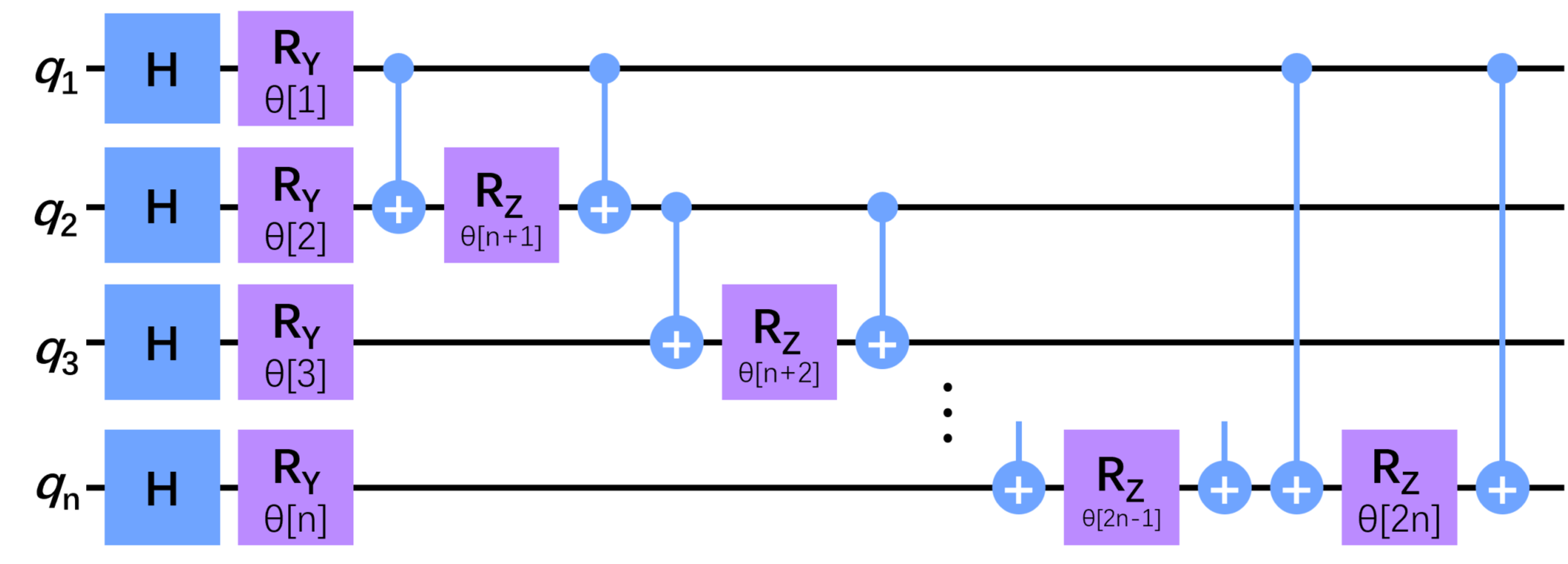}
	}
	\subfigure[Hardware-efficient ansatz]{
		\centering\label{HEAansatz}\includegraphics[height=3.3cm]{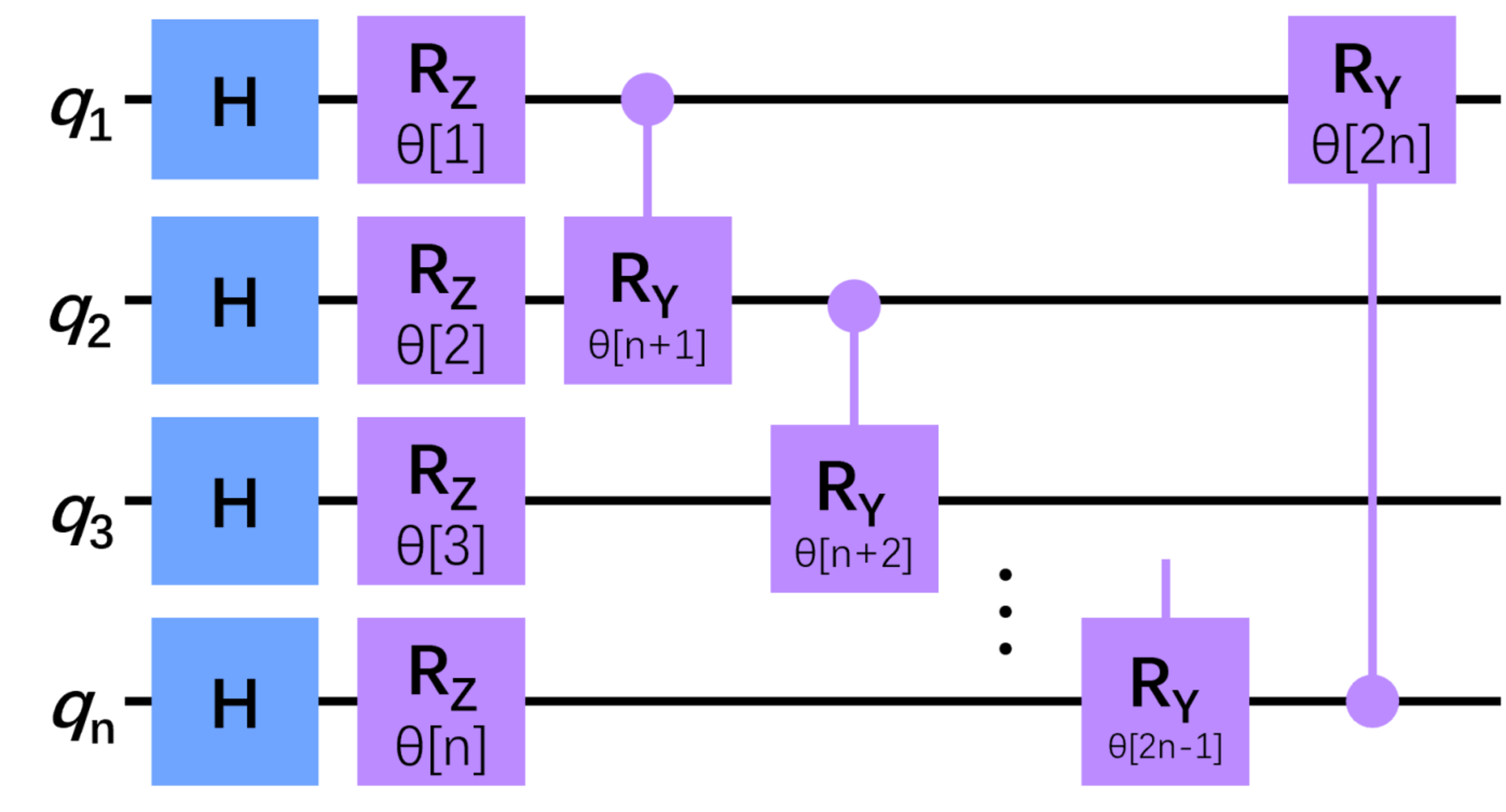}
	}
	\subfigure{
		\centering\label{valid2}\includegraphics[height=3.25cm]{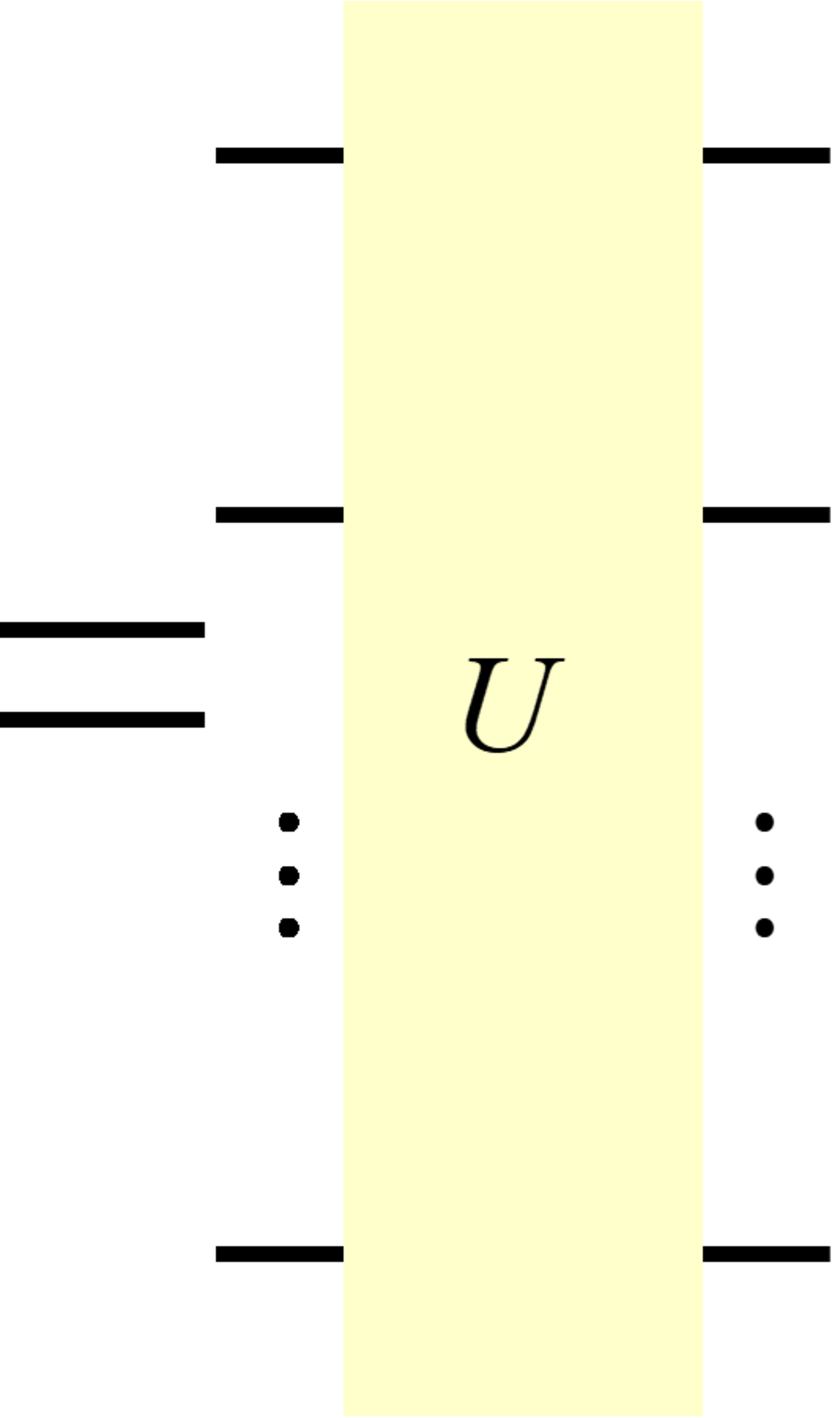}
	}
	\caption{The structure of $U$}\label{f1}
	\end{figure*}

\subsection{Deferred Measurement Principle}\label{DMPS}

The DMP \cite{0.906,0.907,0.91,0.92}, another technical core, is a fundamental result in quantum computing that enables flexible and efficient design of quantum circuits. The rigorous definition of  DMP is as follows.

	A quantum circuit allows for the relocation of a measurement from its intermediate stage to the end of circuit. If the measurement is employed at any point during the circuit, classical control maneuvers can be substituted with conditional quantum operations \cite{0.91}.\label{defDMP}

It asserts that postponing measurements until the end of quantum computing does not impact the probability distribution of the results, allowing measurements to be deferred to more favorable times. In particular, the DMP allows measurements to be commuted with conditional operators, providing significant freedom in designing quantum algorithms. For instance, strategically deferring measurements can reduce the number of required qubits, enable more efficient simulations. Conversely, postponing all measurements until the end of the circuit facilitates pure state analysis of the final output state, enabling the possibility of more precise measurements and error correction. 



\section{Quantum Kernel Self-Attention Mechanism}\label{sec3}

In this section,  the QKSAM theory is introduced to establish a mathematical nexus between QKM and SAM to amalgamate their inherent strengths, thereby laying the cornerstone for the exploration of quantum SAM. QKSAM is defined as follows.
\begin{definition}[QKSAM]
	\begin{equation}\label{QKSAM}
\text{QKSAM}:=|\mathbf{V}\rangle \Theta {{({{\theta }_{4}})}_{|\langle \mathbf{Q}|\mathbf{K}\rangle {{|}^{2}}}}.
	\end{equation} 
\end{definition}
Eq. (\ref{QKSAM}) highly and abstractly recapitulates the below concepts, encompassing Quantum Query State (QQS) $|\mathbf{Q}\rangle$, Quantum Valued State (QVS) $|\mathbf{V}\rangle$, Quantum Key State (QKS) $|\mathbf{K}\rangle$, Quantum Kernel Self-Attention Score (QKSAS) $|\langle \mathbf{Q}|\mathbf{K}\rangle {|}^{2}$, and mathematical connections $\Theta(\theta_{4})$ between quantum registers. The definitions of these concepts are expounded one by one afterwards.


\subsection{Quantum Kernel Self-Attention Score}

 First, the inputs 
 \begin{equation}\label{inputs1}
{{\mathbf{w}}_{i}}=[{{a}_{c}}]_{c=0}^{l-1},{{\mathbf{w}}_{j}}=[{{b}_{c}}]_{c=0}^{l-1}\in {{\mathbb{R}}^{1\times l}}\subseteq \mathcal{X}
 \end{equation}                                          
  in Section \ref{Aa} are embedded into the Hilbert space $\mathcal{F}$ and its dual space $\mathcal{F}^{\dagger}$ respectively, by  QFM and its inverse mapping QFM$^{\dagger}$ in Section \ref{QKMS}, i.e., 
  \begin{equation}\label{map1}
  {{U}_{\phi }}:	{{\mathbf{w}}_{i}}\to |\phi ({{\mathbf{w}}_{i}})\rangle ={{U}_{\phi }}({{\mathbf{w}}_{i}})|\mathbf{0}\rangle_{1} \subseteq \mathcal{F}
  \end{equation} 
  and  
  \begin{equation}\label{map2}
  	{{U}_{\phi }^{\dagger}}: {{\mathbf{w}}_{j}}\to \langle \phi ({{\mathbf{w}}_{j}})|=\langle \mathbf{0}|_{1}U_{\phi }^{\dagger }({{\mathbf{w}}_{j}})\subseteq {{\mathcal{F}}^{\dagger }}.
  \end{equation}
 In Eq. (\ref{inputs1}),  $a_{c}$ and $b_{c}$ are elements in $\mathbf{w}_{i}$ and $\mathbf{w}_{j}$, respectively.  $l$ is the dimension of the element. $\mathcal{X}$ represents the spatial domain where the input data resides. Here, let
   \begin{equation}
n=\left\lceil {{\log }_{2}}l \right\rceil,
 \end{equation} 
 which is used multiple times in the following to denote the number of qubits. In Eq. (\ref{map1}) and Eq. (\ref{map2}), subscript 1 denotes the first quantum register. $\langle \mathbf{0}|_{1}$ is the conjugate transpose of $|\mathbf{0}\rangle_{1}$. $U_{\phi }^{\dagger }$ is the inverse unitary operator of ${{U}_{\phi }}$. As for the explicit structure of ${{U}_{\phi }}$, under limited quantum resources,  the amplitude coding \cite{0.920}
  \begin{equation}
{{U}_{\phi }}:{{\mathbf{w}}_{i}}\to \sum\limits_{c=0}^{l-1}{{{\alpha }_{c}}|c{{\rangle }_{1}}}+\sum\limits_{c=l}^{{{2}^{n}}-1}{0|c{{\rangle_{1} }}}\subseteq \mathcal{F}
  \end{equation} 
  in Fig. \ref{AMPLITUDU} is chosen, where ${{\alpha }_{c}}={{a}_{c}}/\sqrt{\sum\nolimits_{d=0}^{l-1}{a_{d}^{2}}}$ . $a_{c}, a_{d} \in {\mathbf{w}_{i}}$. $l\le {{2}^{n}}$.
   Alternatively, the angle coding \cite{0.921}
  \begin{equation}\label{aaa}
{{U}_{\phi }}:\mathbf{w}_{i}^{'}\to \prod\limits_{c=0}^{n-1}{\left( \underset{d=0}{\mathop{\overset{n-1}{\mathop{\otimes }}\,}}\,{{R}_{e}}(a_{d+c \times n}^{'}) \right)}|\mathbf{0}{{\rangle _{1}}}\subseteq \mathcal{F}
  \end{equation}  
in Fig. \ref{angles} is adopted for simplicity of realization, where $a_{d+c \times n}^{'} \in \mathbf{w}_{i}^{'}$. $\mathbf{w}_{i}^{'}=[a_{c}^{'}]_{c=0}^{{{2}^{n}}-1}$, where $${{a}_{c}^{'}}=\left\{ \begin{array}{*{35}{l}}
	{{a}_{c}}\in {{\mathbf{w}}_{i}} & c\in [0,l-1]  \\
	0 & else  \\
\end{array} \right..$$
$\otimes$ is the tensor product. Since $$e=\left\{ \begin{array}{*{35}{l}} X & c\%3=0   \\	Y & c\%3=1  \\ Z & c\%3=2	 \\ \end{array} \right.,$$ 
${{R}_{e}}\in \{{{R}_{X}},{{R}_{Y}},{{R}_{Z}}\}$, where $R_{X}$, $R_{Y}$ and $R_{Z}$ are  Pauli rotating gates in Tab. \ref{Notations}.

  Subsequently, Eq. (\ref{map1}) and Eq. (\ref{map2}) are subjected to transformations by unitary operator $U(\theta_{1})$ and its inverse operator $U^{\dagger}(\theta_{2})$ respectively, where $\theta_{1}$ and $\theta_{2}$ are trainable parameters, thereby obtaining QQS and QKS corresponding to Eq. (\ref{q1}) and Eq. (\ref{k1})  in Section \ref{Aa}:
  \begin{definition}[QQS]
  	  \begin{equation}\label{QQS}
|{{\mathbf{Q}}}\rangle =U({{\theta }_{1}}){{U}_{\phi }}({{\mathbf{w}}_{i}})|\mathbf{0}\rangle_{1} \subseteq \mathcal{F}.
  	\end{equation}  
  \end{definition}
    \begin{definition}[QKS]
  	\begin{equation}\label{QKS}
\langle \mathbf{K}|=\langle \mathbf{0}|_{1}U_{\phi }^{\dagger }({{\mathbf{w}}_{j}}){{U}^{\dagger }}({{\theta }_{2}})\subseteq {{\mathcal{F}}^{\dagger }}.
  	\end{equation}  
  \end{definition}
  In particular, the structure of the parameter layers $U({{\theta }_{1}})$  in Eq. (\ref{QQS}) is structured by mainstream QAOA ansatz \cite{5.4,5.41} 
  \begin{equation}\label{QAOAAnsatz}
\prod\limits_{c=0}^{n -1}{CNOT[c,f]{{R}_{Z}}[f]CNOT[c,f]}\underset{c=0}{\mathop{\overset{n -1}{\mathop{\otimes }}\,}}\,({{R}_{Y}}[c]H[c])
\end{equation}  
 in Fig.  \ref{QAOAansatz} or hardware-efficient ansatz \cite{5.3,5.31}
  \begin{equation}\label{HEAnsatz}
\prod\limits_{c=0}^{n -1}{C{{R}_{Y}}[c,f]}\underset{c=0}{\mathop{\overset{n -1}{\mathop{\otimes }}\,}}\,({{R}_{Z}}[c]H[c])
  \end{equation}  
  in Fig. \ref{HEAansatz}, where  $$f=\left\{ \begin{array}{*{35}{l}}
  	c+1 & c\ne n-1  \\
  	0 & c=n -1  \\
  \end{array} \right..$$
	In Eq. (\ref{QAOAAnsatz}) and Eq. (\ref{HEAnsatz}),  $CNOT$, $H$ and $C{R}_{Y}$ stand for CNOT gates, Hadamard gates and controlled Y gates in Tab. \ref{Notations}, respectively.  In addition, quantum coordinates \cite{1.0101}  are employed to delineate the two $U$ structures in Fig. \ref{f1}. Specifically, for the two-element bracket $[c,f]$, $c$ and $f$ indicate the control and result bit positions. For the single-element bracket $[c]$ without  control bit, $c$ designates the position of result bit. Analogous to the idea of Eq. (\ref{QQS}), QVS can be defined:
	\begin{definition}[QVS]
		\begin{equation}\label{QVS}
			|\mathbf{V}\rangle =U({{\theta }_{3}}){{U}_{\phi }}({{\mathbf{w}}_{j}})|\mathbf{0}{{\rangle_{2} }}\subseteq \mathcal{F}.
		\end{equation}  
	\end{definition}
	$\theta_{3}$ in Eq. (\ref{QVS}) is the trainable parameter. Notably, since the subscript of Eq. (\ref{QVS}) is 2, QVS is located in the second quantum register. 

	Based on the explicit definitions of QSS and QKS above, it is easily related to Eq. (\ref{QKM1}), which means that  the QKSAS is obtained according to Fig. \ref{QKM2} in Section \ref{QKMS}:
	\begin{definition}[QKSAS]
		  \begin{equation}\label{QKSAS}
		  	\begin{aligned}
	 |\langle \mathbf{Q}|\mathbf{K}\rangle {{|}^{2}}&=|{{\langle \mathbf{0}{{|}_{1}}U_{\phi }^{\dagger }({{\mathbf{w}}_{j}}){{U}^{\dagger }}({{\theta }_{2}})U({{\theta }_{1}}){{U}_{\phi }}({{\mathbf{w}}_{i}})|\mathbf{0}\rangle _{1}}}{{|}^{2}} \\ 
	& \ge 0 .\\ 
\end{aligned}
	\end{equation}  
	\end{definition}

	\begin{proof} Let $S({{\mathbf{w}}_{j}},{{\mathbf{w}}_{i}},{{\theta }_{1}},{{\theta }_{2}})=U_{\phi }^{\dagger }({{\mathbf{w}}_{j}}){{U}^{\dagger }}({{\theta }_{2}})U({{\theta }_{1}}){{U}_{\phi }}({{\mathbf{w}}_{i}})$ and $S^{\dagger}({{\mathbf{w}}_{j}},{{\mathbf{w}}_{i}},{{\theta }_{1}},{{\theta }_{2}})$ be the complex conjugate of $S({{\mathbf{w}}_{j}},{{\mathbf{w}}_{i}},{{\theta }_{1}},{{\theta }_{2}})$. 
		In accordance with Fig. \ref{adjoint}, if using the projection operator $P=|\mathbf{0}\rangle_{1}\langle\mathbf{0}|_{1}$ , the ultimate outcome of the first quantum register can be measured as $|\langle \mathbf{Q}|\mathbf{K}\rangle {{|}^{2}}={{\langle \mathbf{0}{{|}_{1}}S({{\mathbf{w}}_{j}},{{\mathbf{w}}_{i}},{{\theta }_{1}},{{\theta }_{2}})P{{S}^{\dagger }}({{\mathbf{w}}_{j}},{{\mathbf{w}}_{i}},{{\theta }_{1}},{{\theta }_{2}})|\mathbf{0}\rangle }_{1}} ={{\langle \mathbf{0}{{|}_{1}}S({{\mathbf{w}}_{j}},{{\mathbf{w}}_{i}},{{\theta }_{1}},{{\theta }_{2}})|\mathbf{0}\rangle }_{1}}{{\langle \mathbf{0}{{|}_{1}}{{S}^{\dagger }}({{\mathbf{w}}_{j}},{{\mathbf{w}}_{i}},{{\theta }_{1}},{{\theta }_{2}})|\mathbf{0}\rangle }_{1}}$.
Since ${{({{\langle \mathbf{0}{{|}_{1}}S({{\mathbf{w}}_{j}},{{\mathbf{w}}_{i}},{{\theta }_{1}},{{\theta }_{2}})|\mathbf{0}\rangle_{1}} })}^{\dagger }}={{\langle \mathbf{0}{{|}_{1}}{{S}^{\dagger }}({{\mathbf{w}}_{j}},{{\mathbf{w}}_{i}},{{\theta }_{1}},{{\theta }_{2}})|\mathbf{0}\rangle_{1} }}$, $|\langle \mathbf{Q}|\mathbf{K}\rangle {{|}^{2}}$ is the square of the module of ${{{{\langle \mathbf{0}{{|}_{1}}S({{\mathbf{w}}_{j}},{{\mathbf{w}}_{i}},{{\theta }_{1}},{{\theta }_{2}})|\mathbf{0}\rangle_{1}} }}}$,  thus giving Eq. (\ref{QKSAS}).   Currently, only $P$ is taken into account. If each basis state is examined, the attention distribution can be obtained as a probability vector.\null\hfill\ $\square$
	\end{proof}

\subsection{Mathematical Link between Quantum Registers}	
So far, the definitions of QSS, QKS, QVS and QKSAS, as well as the specific structures of ${{U}_{\phi }}$  and $U$, have been clarified, which suggests that the quantum counterparts of Eq. (\ref{q1}), Eq. (\ref{k1}), Eq. (\ref{v1}) and Eq. (\ref{weight}) in Section \ref{Aa} have been created on a quantum computer. First, Eq. (\ref{QQS}), Eq. (\ref{QKS}) and Eq. (\ref{QVS}) are briefly reviewed here. Eq. (\ref{QQS}) and Eq. (\ref{QVS}) are formally identical, while Eq. (\ref{QKS}) is the conjugate transposed form of Eq. (\ref{QQS}). It is worth emphasizing that the parameters in Eq. (\ref{QQS}), Eq. (\ref{QKS}) and Eq. (\ref{QVS}) may not necessarily coincide. Another difference is that Eq. (\ref{QVS}) is not in the same quantum register as Eq. (\ref{QQS}) (or Eq. (\ref{QKS})).  On the basis of the above, next, a mathematical link $\Theta(\theta_{4})$ between the first and second quantum registers is constructed using DMP. More precisely, this mathematical link $\Theta(\theta_{4})$ designates the measurement outcome QKSAS as the control signal for the second quantum register:
\begin{definition}[Mathematical Link $\Theta(\theta_{4})$]
	The mathematical link $\Theta(\theta_{4})$ as shown in Fig. \ref{CUDMP} specifically means that according to the DMP, when the first quantum register measures the ground state $|\mathbf{0}\rangle$ with the probability of QKSAS, then the unitary operator
	  	  \begin{equation}\label{QQ121}
C{{U}_{\text{DMP}}}({{\theta }_{4}})=\underset{c=0}{\overset{n -1}{\mathop \otimes }}\,(R_{X}[c]C{{R}_{Y}}[c,c+n])
	\end{equation}  containing the trainable parameter $\theta_{4}$  is applied to the second quantum register.
\end{definition}
In Eq. (\ref{QQ121}), $CR_{Y}$ spans the first and second quantum registers, so its quantum coordinate is $[c,c+n ]$, where $n$ is the interval between the control and result bits. In contrast, $R_{X}$ which includes only the result bit, is confined solely to the first quantum register. Now, it is necessary to show that the final result is not changed by the mathematical link $\Theta(\theta_{4})$.

\begin{proof}	The quantum state before the first register is measured in Fig. \ref{CUDMP} is noted as $|\Phi \rangle $ which can be re-represented as a superposition of orthogonal basis vectors, that is  $|\Phi \rangle =\sum\nolimits_{i=0}^{{{2}^{n}}-1}{{{c}_{i}}|i\rangle }$.
Here, $\sum\nolimits_{i=0}^{{{2}^{n}}-1}{c_{i}^{2}}=1$. Since the controlled operators $C{{U}_{\text{DMP}}}$ places the unitary operators $U_{\text{DMP}}$ on the second quantum register only when the controlled qubits are $|\mathbf{0}\rangle_{1}$, $|\Phi \rangle $ is again denoted as 
	$|\Phi \rangle ={{c}_{0}}|\mathbf{0}\rangle_{1} +{{\tilde{c}}_{1}}|\mathbf{\tilde{1}}\rangle_{1} $,
where	${{\tilde{c}}_{1}}|\mathbf{\tilde{1}}\rangle _{1}=\sum\nolimits_{i=1}^{{{2}^{n}}-1}{{{c}_{i}}|i\rangle_{1} }$. In this way, the quantum state $|\mathbf{0}\rangle_{1}$ is prominently listed. Besides, due to the presence of entangled quantum gates, exemplified by CNOT gates, the comprehensive quantum state can be designated as 	
$|\Phi \rangle ={{c}_{0}}|\mathbf{0}{{\rangle _{1}}}|{{\Phi }^{+}}{{\rangle _{2}}}+{{\tilde{c}}_{1}}|\mathbf{\tilde{1}}{{\rangle_{1} }}|{{\Phi }^{-}}{{\rangle_{2} }}$.

\begin{example}[Quantum Control]
	From the quantum control point of view, $|\Phi \rangle $ becomes 
	$C{{U}_{\text{DMP}}}|\Phi \rangle ={{c}_{0}}|\mathbf{0}{{\rangle_{1} }}{{U}_{\text{DMP}}}|{{\Phi }^{+}}{{\rangle _{2}}}+{{\tilde{c}}_{1}}|\mathbf{\tilde{1}}{{\rangle_{1} }}|{{\Phi }^{-}}{{\rangle _{2}}}
	$
	after being applied the controlled quantum gates ${U}_{\text{DMP}}$.  If the base state $|\mathbf{0}\rangle_{1} $  is measured, then the post-measurement state is
\begin{equation}\label{DDQ}
\begin{aligned}
	& \text{   }({{P}_{0}}\otimes {{I}_{n}})C{{U}_{\text{DMP}}}|\Phi \rangle  \\ 
	& ={{P}_{0}}{{c}_{0}}|\mathbf{0}{{\rangle_{1} }}{{U}_{\text{DMP}}}|{{\Phi }^{+}}{{\rangle _{2}}}+{{P}_{0}}{{{\tilde{c}}}_{1}}|\mathbf{\tilde{1}}{{\rangle_{1} }}|{{\Phi }^{-}}{{\rangle_{2} }} \\ 
	& =c_{0}|\mathbf{0}{{\rangle_{1} }}U_{\text{DMP}}|{{\Phi }^{+}}{{\rangle_{2} }},   \\ 
\end{aligned}
\end{equation}
where ${P}_{0}=|\mathbf{0}\rangle_{1}\langle \mathbf{0}|_{1}$ is the projection operator. $I_n$ denotes the identity matrix acting on the second quantum register. 
\end{example}

	\begin{example} [DMP]
		Now another case i.e. the case of intermediate measurements followed by the imposition of unitary operator $U_{\text{DMP}}$ is considered. Similarly, if the ground state $|\mathbf{0}\rangle_{1} $  is measured, the state after the measurement is
			$P_0|\Phi\rangle =  c_{0}|\mathbf{0}{{\rangle_{1} }}|{{\Phi }^{+}}{{\rangle_{2} }} $.
		The unitary operator $U_{\text{DMP}}$ is then applied to the second register conditional on the appearance of $P_0|\Phi\rangle$:
		\begin{equation}\label{PS2}
			c_0(I_n\otimes U_{\text{DMP}})|\mathbf{0}{{\rangle_{1} }}|{{\Phi }^{+}}{{\rangle _{2}}}=c_{0}|\mathbf{0}{{\rangle _{1}}}U_{\text{DMP}}|{{\Phi }^{+}}{{\rangle _{2}}}.
		\end{equation}
	\end{example}
	
	
	Comparing Eq. (\ref{DDQ}) with Eq. (\ref{PS2}), it is evident that the post-measurement states are identical in both scenarios, irrespective of the measurement outcome. Since the measurement outcome probability is the squared norm of the  post-measurement state, it follows that both measurement outcomes have equal probabilities in each scenario.  Thus the action of the DMP does not affect the final result. \null\hfill\ $\square$
	
\end{proof}

Eventually, the complete QKSAM can be interpreted. Eq. (\ref{QKSAM}) covers all preceding crucial definitions in a highly abstract manner. It indicates that the link $\Theta$ between the measured ground state $|\mathbf{0}\rangle$ in the first quantum register and the QVS in the second quantum register is determined by the QKSAS that depends on Eq. (\ref{QQS}) and Eq. (\ref{QKS}).

Overall, certain crucial concepts in QKSAM are inherited from SAM, including the quantum variants of query vectors, key vectors, and value vectors, denoted as QQS, QKS, and QVS. However, please note that QKSAM is not an imitator of SAM. Therefore, the divergences between QKSAM and SAM, seen as a mechanism innovation,  should be paid more attention. First of all, QKSAM model is realized by quantum ansatzes deployed on a quantum computer, exemplified by the use of amplitude coding and the QAOA ansatz as described in this paper. Second, the QKSAS of Eq. (\ref{QQS}) and Eq. (\ref{QKS}) are estimated with QKM after projecting the classical data into Hilbert space. At last, the link between Eq. (\ref{weight}) and Eq. (\ref{v1}) in Section \ref{Aa} is reshaped into a DMP-based control relation $\Theta$, thus avoiding wasting a large number of qubits to construct a large multiplicative network under the current constraints of quantum resources, making the QKSAM scheme more feasible and cost-effective. 


\section{Quantum Kernel Self-Attention Network}\label{workflow}

\begin{figure*}
	
	\begin{minipage}[t]{0.43\textwidth}
		\centering
		\includegraphics[height=5.2cm]{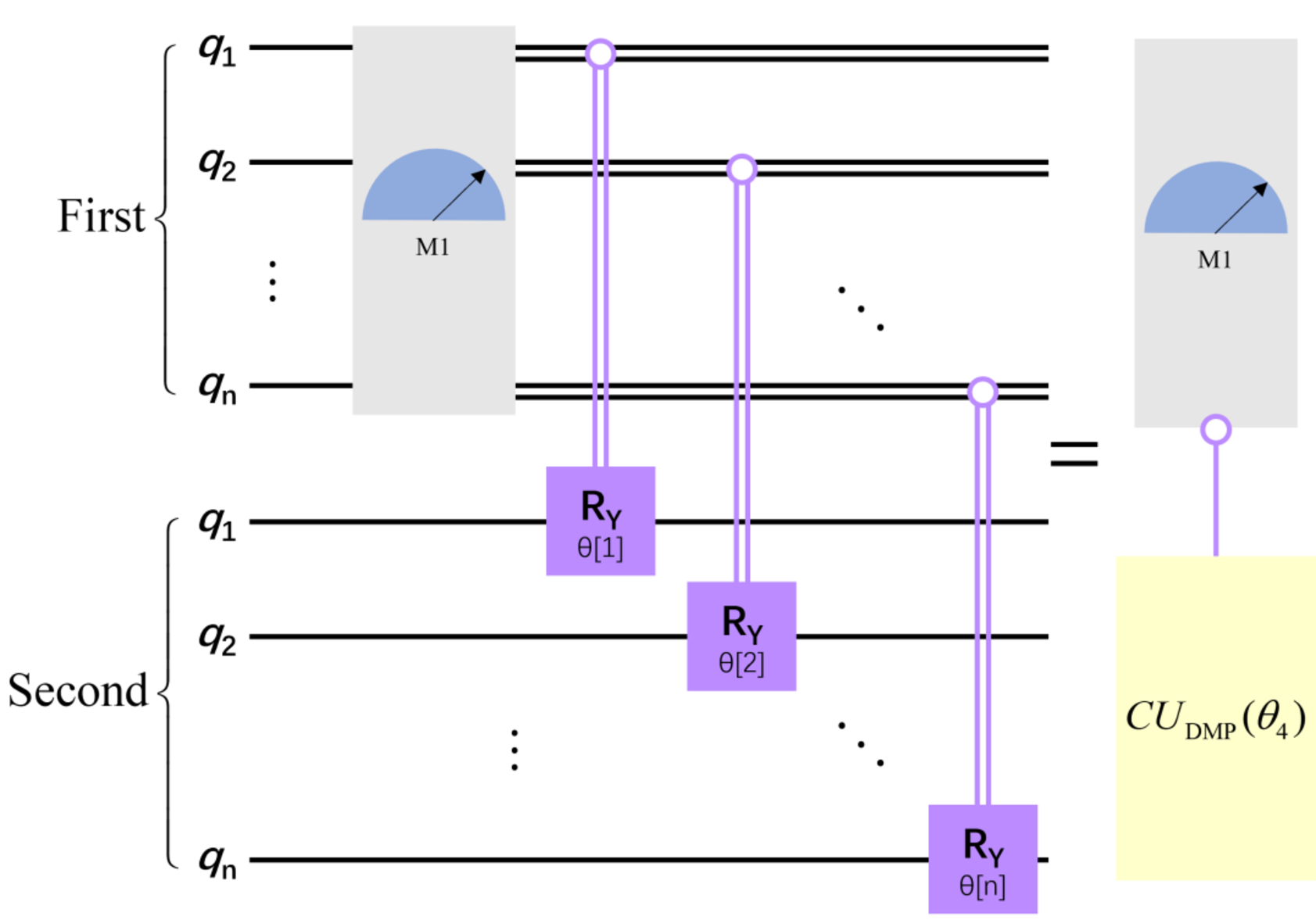}
		\caption{The mathematical link $\Theta $}\label{CUDMP}
	\end{minipage}
	\begin{minipage}[t]{0.6\textwidth}
		\centering
		\includegraphics[height=5.2cm]{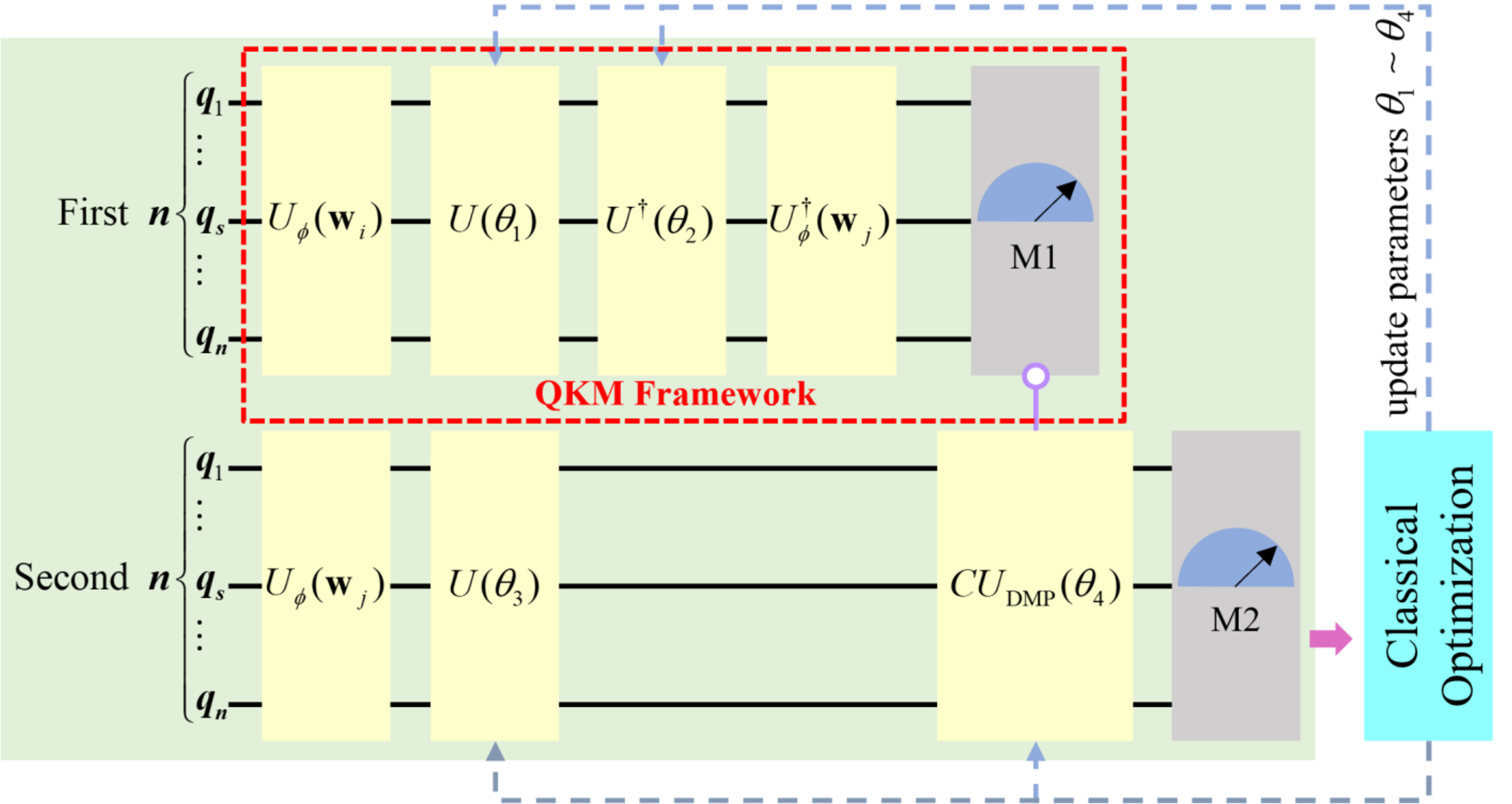}
		\caption{The framework of QKSAN}\label{QKSAN}
	\end{minipage}
\end{figure*}

In this section, the QKSAM-based QKSAN framework and its workflow are explained in detail. The QKSAN framework in Fig. \ref{QKSAN} is designed as a hybrid quantum-classical architecture.

The QKSAM ansatz framework on a quantum computer, depicted in the green box, is horizontally divided into two registers, each with $n$ qubits. Among them, the first register is inspired by the structure of Fig. \ref{adjoint} for solving QKSAS in Eq. (\ref{QKSAS}). Furthermore, the structures of ${{U}_{\phi }}$ and $U$ can be found in Fig. \ref{f0} and Fig. \ref{f1}, with $U_{\phi }^{\dagger }$ and ${{U}^{\dagger }}$ as their respective conjugate transpose operators. The second register is mainly used to prepare QVS in Eq. (\ref{QVS}) as well as to generate the final output M2. The mathematical link $\Theta(\theta_{4})$  between the first and second registers is represented by the first register output M1 and ansatz $C{{U}_{\text{DMP}}}({{\theta }_{4}})$ in Fig. \ref{QKSAN}. 

The elements situated beyond the confines of the green box pertain to the classical optimization realm. In this domain, a classical optimizer is enlisted, exemplified by the quantum natural simultaneous perturbation stochastic approximation optimizer \cite{5.0} or the quantum natural gradient optimizer \cite{5.01}. These optimizers are harnessed iteratively to adjust the parameters until convergence of the loss function is attained. Notably, this paper employs the Nesterov momentum optimizer \cite{5.02}, a variant that integrates a momentum component with the gradient descent, thereby considering historical gradients in the optimization process.

Based on the above framework, its workflow is described here.

Step 1: The initial quantum state
\begin{equation}\label{Step1}
	|\Psi {{\rangle_{1} }}\otimes |\Phi {{\rangle _{2}}}=|\mathbf{0}\rangle_{1} \otimes |\mathbf{0}\rangle_{2}
\end{equation}
is prepared in the first and second quantum registers.

Step 2: ${{U}_{\phi }}({{\mathbf{w}}_{i}})$, $U({{\theta }_{1}})$, ${{U}^{\dagger }}({{\theta }_{2}})$ and ${U}^{\dagger}_{\phi }({{\mathbf{w}}_{j}})$ are applied in order for the first register while ${U}_{\phi }({{\mathbf{w}}_{j}})$ and $U({{\theta }_{3}})$ are used in turn for the second one, i.e.
\begin{equation}\label{Step2}
\begin{aligned}
	& |\Psi {{\rangle _{1}}}\otimes |\Phi {{\rangle_{2} }}\xrightarrow{{{U}_{\phi }}({{\mathbf{w}}_{i}})\otimes {{U}_{\phi }}({{\mathbf{w}}_{j}})}{{({{U}_{\phi }}({{\mathbf{w}}_{i}})|\mathbf{0}\rangle_{1} }})\otimes {{({{U}_{\phi }}({{\mathbf{w}}_{j}})|\mathbf{0}\rangle_{2} }}) \\ 
	& \xrightarrow{U({{\theta_{1} }})\otimes U({{\theta_{3} }})}{{(U({{\theta _{1}}}){{U}_{\phi }}({{\mathbf{w}}_{i}})|\mathbf{0}\rangle_{1} }})\otimes {{(U({{\theta_{3} }}){{U}_{\phi }}({{\mathbf{w}}_{j}})|\mathbf{0}\rangle_{2} }}) \\ 
	& \xrightarrow{{{U}^{\dagger }}({{\theta_{2} }})\otimes I}{{({{U}^{\dagger }}({{\theta _{2}}})U({{\theta_{1} }}){{U}_{\phi }}({{\mathbf{w}}_{i}})|\mathbf{0}\rangle_{1} }})\otimes {{(IU({{\theta_{3} }}){{U}_{\phi }}({{\mathbf{w}}_{j}})|\mathbf{0}\rangle_{2} }}) \\ 
	& \xrightarrow{U_{\phi }^{\dagger }({{\mathbf{w}}_{j}})\otimes I}\begin{array}{*{35}{l}}
		{{(U_{\phi }^{\dagger }({{\mathbf{w}}_{j}}){{U}^{\dagger }}({{\theta _{2}}})U({{\theta_{1} }}){{U}_{\phi }}({{\mathbf{w}}_{i}})|\mathbf{0}\rangle_{1} }})  \\
		\otimes {{(I\times IU({{\theta_{3} }}){{U}_{\phi }}({{\mathbf{w}}_{j}})|\mathbf{0}\rangle_{2} }})  \\
	\end{array} ,\\ 
\end{aligned}
\end{equation}
where $I$ is an identity matrix of size $2^n \times 2^n$.

\begin{figure*}[h]
	\centering
	\includegraphics[scale=0.18]{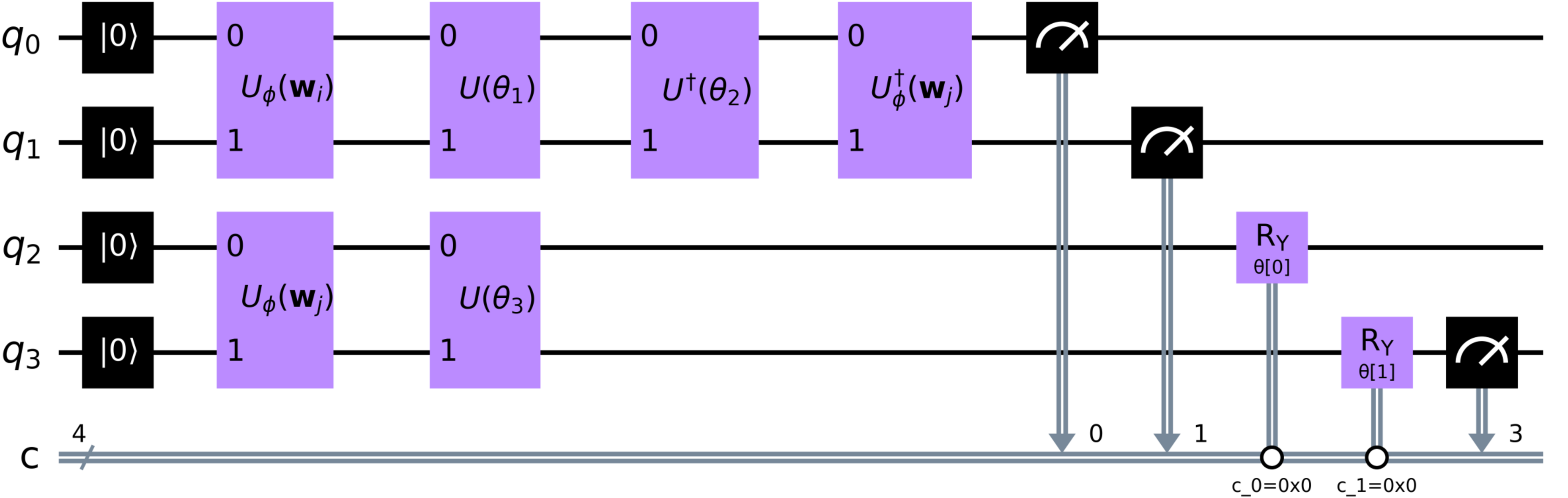}
	\caption{A 4-qubit QKSAN drawn by IBM Qiskit}\label{ansatz}
\end{figure*}

Step 3: DMP is applied in the middle of the quantum ansatz of QKSAN, i.e.
\begin{equation}
	|\Phi {{\rangle _{2}}}=\left\{ \begin{array}{*{35}{l}}
		C{{U}_{\text{DMP}}}({{\theta _{4}}})T, & P(|\mathbf{0}\rangle_{1} )=\text{QKSAS}  \\
		T, & else  \\
	\end{array} \right.,
\end{equation}
where $T=U({{\theta }_{3}}){{U}_{\phi }}({{\mathbf{w}}_{j}})|\mathbf{0}{{\rangle_{2} }}$. In this step, the first quantum register implements the measurement M1 in Fig. \ref{QKSAN} in the middle of the circuit according to the DMP to solve for  QKSAS. In other words, the first quantum register is waiting for the occurrence  of the ground state $|\mathbf{0}\rangle_{1}$  with a probability corresponding to QKSAS. Once the ground state $|\mathbf{0}\rangle_{1}$ is present, $C{{U}_{\text{DMP}}}$ is applied in the second quantum register, otherwise no operation is performed on this  quantum register. Ultimately, measurement M1 induces the collapse of quantum state $|\Psi {{\rangle _{1}}}$ in the first register, preserving only quantum state $|\Phi {{\rangle_{2} }}$ in the second register. Totally, the incorporation of DMP holds great significance as it efficiently slashes the quantum resource demand for QKSAN in computation by half, which is an invaluable advantage, particularly in the present era of constrained quantum resources. Simultaneously, it amplifies the adaptability and feasibility of implementing QKSAN.

Step 4: The expectation value 
\begin{equation}
	\mathbb{E}={{\langle \Phi {{|}_{2}}P|\Phi \rangle _{2}}}
\end{equation} 
is obtained from the measurement M2 in Fig. \ref{QKSAN}, where $P$ is the projection operator.

Step 5: Cost function is
\begin{equation}
	f(\mathcal{D},\mathcal{D}',{{\boldsymbol{\theta }}})=\frac{1}{m}\sum\limits_{i=1}^{m}{{{[{{y}_{i}}-\text{sgn} (\mathbb{E})]}^{2}}},
\end{equation} 
where $\text{sgn}(\cdot)$ is the sign function. $m$ is the number of terms. ${y}_{i}$ stands for real label. The parameter ${{\boldsymbol{\theta }}}$ contains ${{{\theta }}_{1}}$ to ${{{\theta }}_{4}}$. The optimization rules for the Nesterov momentum optimizer are as follows:
\begin{equation}
	\left\{ \begin{aligned}
		& {{\boldsymbol{\theta }}^{(t+1)}}={{\boldsymbol{\theta }}^{(t)}}-{{a}^{(t+1)}} \\ 
		& {{a}^{(t+1)}}=\gamma \cdot {{a}^{(t)}}+\eta \cdot \nabla f(\mathcal{D},{{\mathcal{D}}^{\prime }},{\boldsymbol{\theta }^{(t)}}-\gamma \cdot {{a}^{(t)}}) \\ 
	\end{aligned} \right.\,
\end{equation} 
where the superscript $t$ represents the $t$-th iteration and $t+1$ stands for the next iteration. The momentum term $\gamma$ is adjustable and generally takes the value 0.9. $\eta$ is the learning rate. $a^{(t+1)}$ and $a^{(t)}$ are accumulator terms. $\nabla f(\mathcal{D},{{\mathcal{D}}^{\prime }},{\boldsymbol{\theta }^{(t)}}-\gamma \cdot {{a}^{(t)}})$ denotes the gradient.

In the end, all the above steps are repeated until the cost function converge. 

To sum up, by analyzing Step 1 to 5 above, the operation mechanism of QKSAN in the classification problem is revealed, which lays a theoretical foundation for the experiment.

\section{Experiment}\label{sec4}

In this section, a comprehensive assessment of the QKSAN framework for MNIST and Fashion MNIST binary classification tasks is conducted via the PennyLane and IBM Qiskit platforms. The experiments consist of four parts:
\begin{itemize} 
	\item The performance of four QKSAN sub-models and QSAN \cite{0.901} is evaluated and compared in MNIST and Fashion MNIST binary classification task in an ideal environment with no circuit noise and the same classical optimizer configuration.
	\item QKSAS with exponential characterization capability generated by AmHE model is demonstrated.
	\item To evaluate the performance of QKSAN in a real quantum computer, the effects of bit flip error and amplitude damping error  \cite{1.01} on AmHE are inspected. 
	\item A inductive comparison is made between QKSAN and the same similar schemes  \cite{0.900,0.901,0.902}.
\end{itemize}

\subsection{Models}

Recall that in Section \ref{sec3}, quantum amplitude coding or quantum angle coding in Fig. \ref{f0} is suggested for the structure of the data embedding layer ${{U}_{\phi }}$, while hardware efficient ansatz or QAOA ansatz in Fig. \ref{f1} is chosen for the structure of the trainable layer ${{U}}$. However, it is not yet foreseeable which combination of a coding  method and a trainable layer would lead to a relatively optimal classification result for QKSAN. Therefore, the two coding methods and two trainable layer structures mentioned above are paired separately, resulting in four structures, namely AmHE, AnHE, AmQAOA, and AnQAOA, which are subsequently validated one by one. After confirming the specific models, it is imperative to consider the limited availability of public quantum resources, exemplified by provision of only six to seven  free qubits from IBM. For this reason, a four-qubit experiment is showcased in this section to ensure that the quantum resource is within seven qubits. More specifically, a QKSAN model, created by IBM Qiskit platform, is depicted in Fig. \ref{ansatz}, with four qubits denoted as $q_0$ to $q_3$.  $c$ in Fig. \ref{ansatz} denotes the classical register used to store the measurement results. Finally, the measurement on $q_3$ is regarded as predictive labels for binary classification.  In short, the concrete model of Fig. \ref{ansatz} is fully compliant with the architecture in Fig. \ref{QKSAN} and the workflow in Subsection \ref{workflow}, and is employed as a classification model for MNIST and Fashion MNIST.

\subsection{Dataset} 

MNIST \cite{5.7} and Fashion MNIST \cite{5.8} are well-established and broadly adopted benchmark datasets in the realm of machine learning, serving as cornerstones for numerous studies in the field. They both consist of 10,000 test images and 60,000 training images, each containing $28\times28$ pixel points. For our experimentation, 550 images labeled 0 and another 550 with label 1 are chosen from MNIST (or Fashion MNIST) as the dataset. From the pool of 550 images for each label, 500 images are randomly picked as the training set, while the remaining images are served as the test set. Moreover, to cater to the aforementioned four models, the dimensionality of all images in the dataset has been compressed to 8 dimensions using the principal component analysis algorithm. 
Obviously, severe image compression results in the loss of a significant amount of critical information, thereby presenting challenges in image classification.

\begin{table*}[h]
	\small 
	\centering
	\def\tablename{Tab.}
	\caption{Experimental configuration}
	\label{Experimental comparison}
	\begin{tabular}{lccccc}
		
		\toprule
		\multirow{2}{*}{Indicators} & \multicolumn{5}{c}{Models} \\
		\cmidrule(lr){2-6}
		& AmHE & AnHE & AmQAOA & AnQAOA  & QSAN \cite{0.901}\\
		\midrule
		parameters    & 11                        & 11                    & 11               & 11   &9      \\ 
		
		layers        & 6                         & 5                    & 7               & 6      & 26   \rule{0pt}{12pt} \\  
		
		qubits & \multicolumn{4}{c}{4}                  &8                                  \rule{0pt}{12pt}    \\
		
		learning rate & \multicolumn{5}{c}{0.09}                                               \rule{0pt}{12pt}         \\
		
		loss function & \multicolumn{5}{c}{square loss}                                            \rule{0pt}{12pt}            \\
		
		batch\_size   & \multicolumn{5}{c}{30}                                                 \rule{0pt}{12pt}          \\
		
		step          & \multicolumn{5}{c}{120 steps}                                               \rule{0pt}{12pt}          \\
		
		optimizer     & \multicolumn{5}{c}{Nesterov Momentum Optimizer: $\gamma = 0.9$}              \rule{0pt}{12pt}                       \\
		\bottomrule
	\end{tabular}
\end{table*}
\begin{figure}[h]
	\centering	
	\centering\includegraphics[scale=0.28]{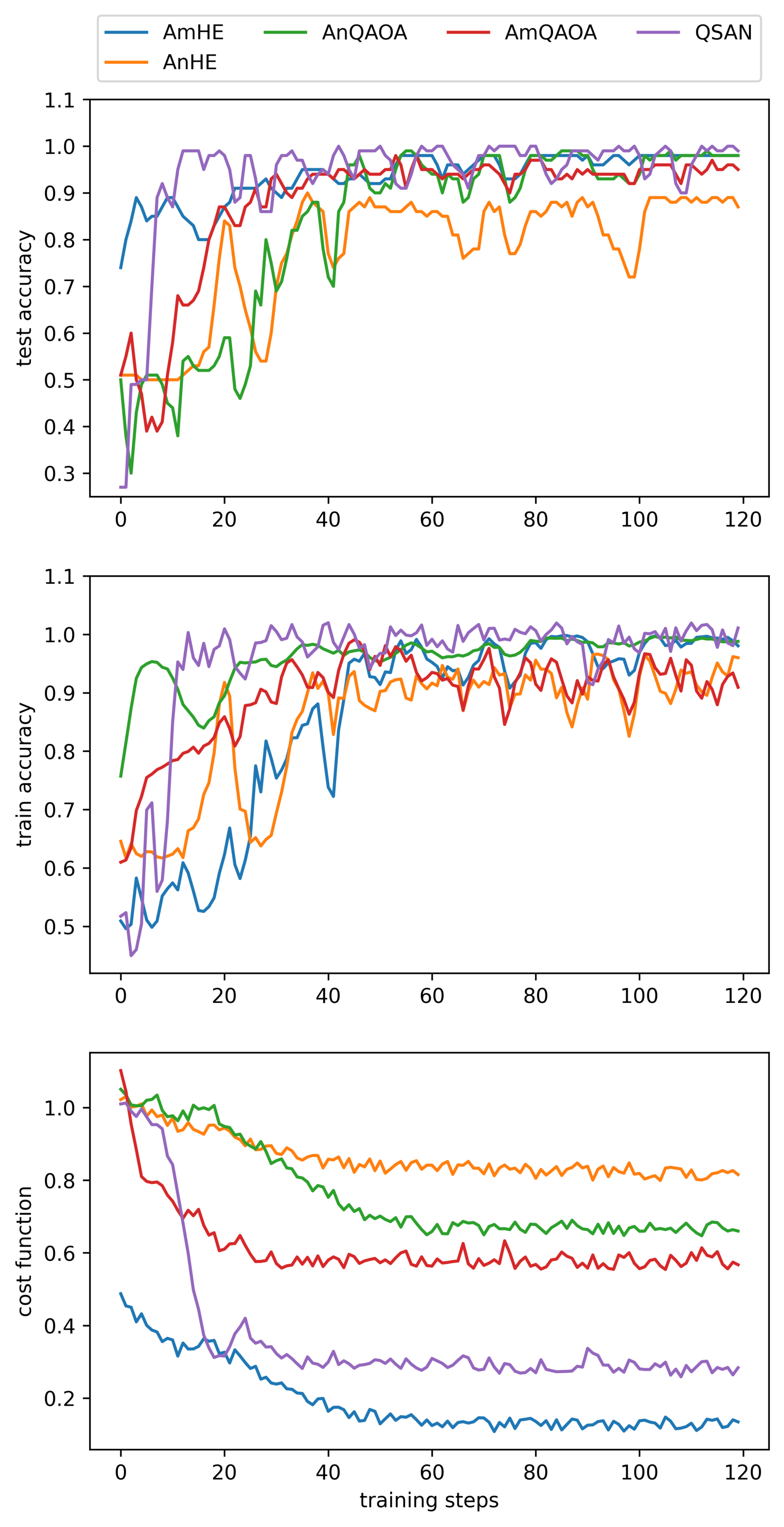}
	\caption{Results of binary classification of MNIST dataset}
	\label{MNIST}
\end{figure}
\begin{figure}[h]
	\centering	
	\centering\includegraphics[scale=0.28]{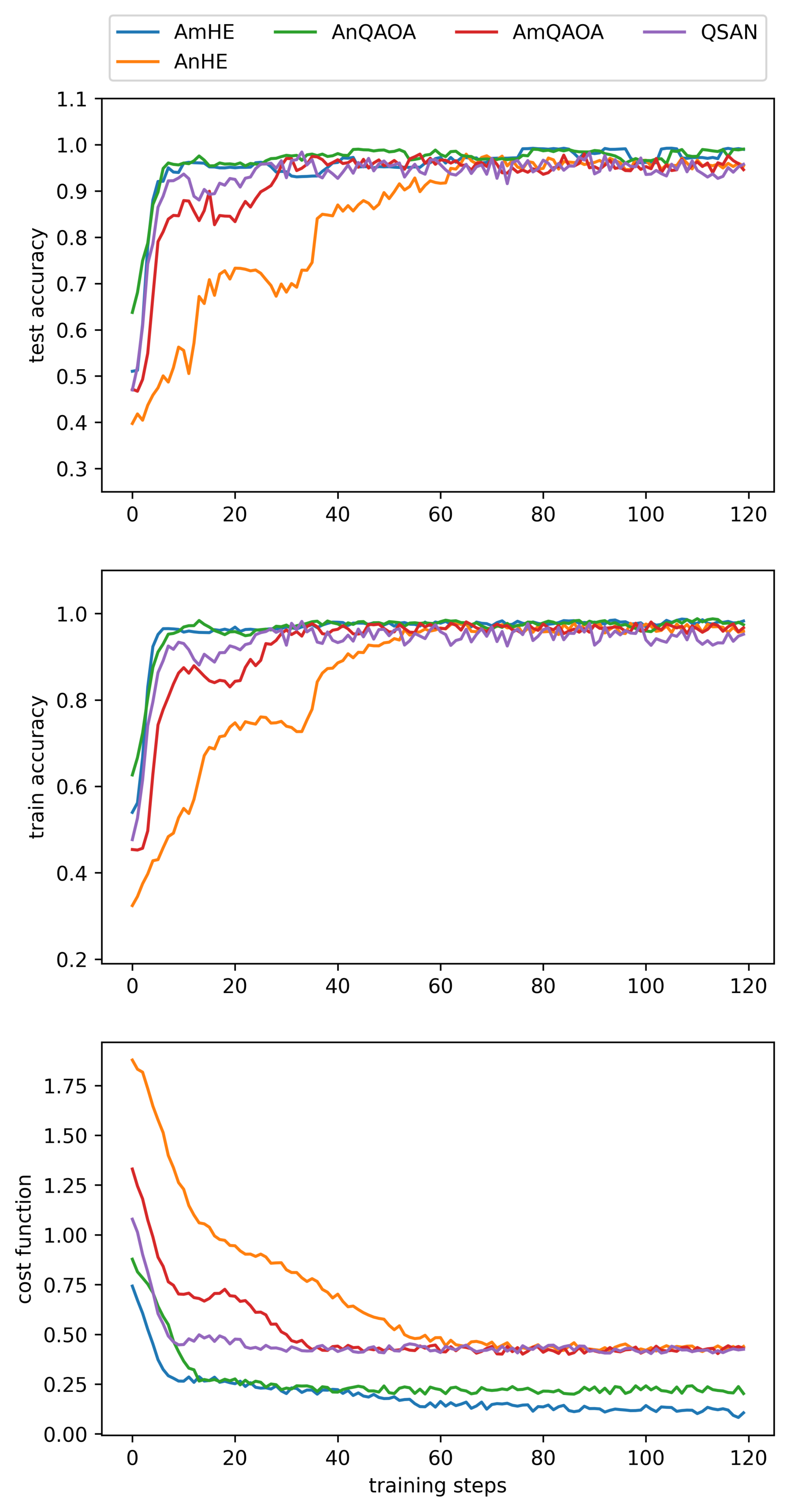}
	\caption{Results of binary classification of Fashion MNIST}
	\label{lr}
\end{figure}
\begin{figure}[h]
	\centering\includegraphics[scale=0.34]{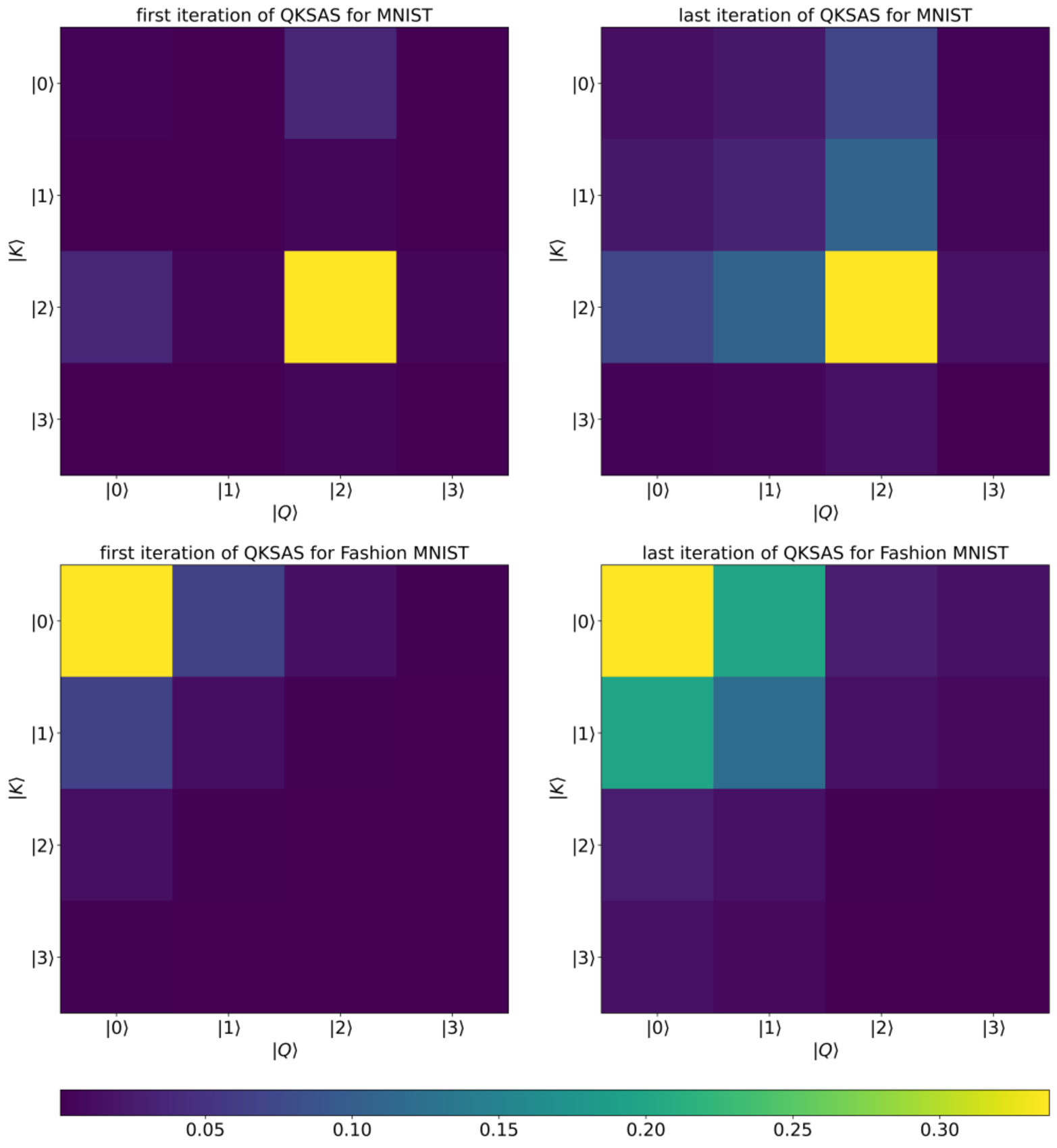}
	\caption{QKSAS of AmHE}
	\label{QAttenscore}
\end{figure}
\begin{figure}[h]
	\centering\includegraphics[scale=0.28]{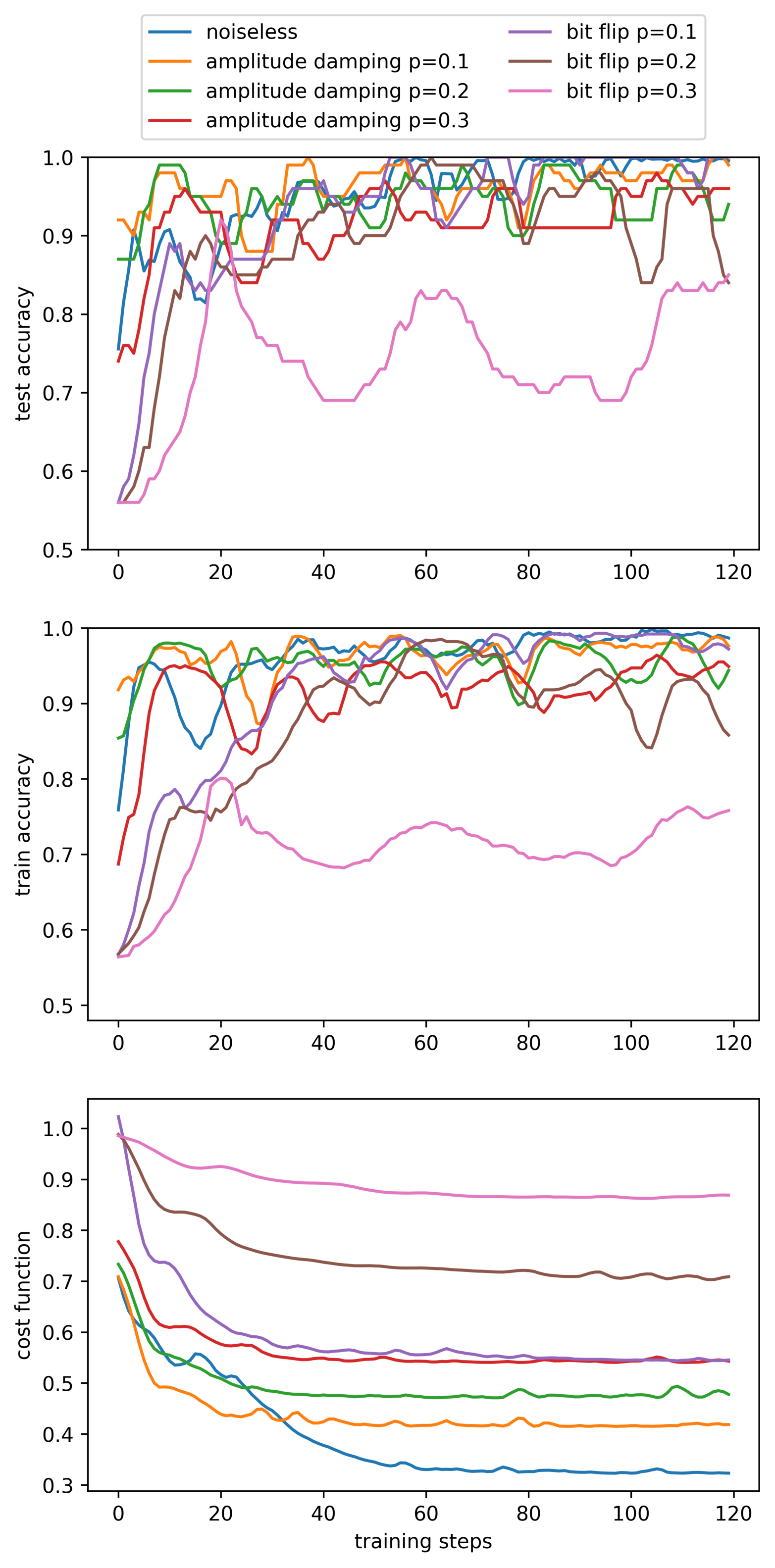}
	\caption{Noise experiments for AmHE}
	\label{Noise_experiments}
\end{figure}
\begin{table*}[h]
	\small
	\def\tablename{Tab.}
	\centering
	\caption{Method Comparison}
	\label{Method Comparison}
		
		\begin{tabular}{lcccc}
			
			\toprule
			\multirow{2}{*}{Indicators} & \multicolumn{4}{c}{Models}                                           \\ \cmidrule(lr){2-5} 
			&QKSAN & Niu’s \cite{0.900}& Zhao’s \cite{0.901}& Li’s \cite{0.902}    \\ \midrule
			realizability on quantum computers & completely & $\times$ & completely & partially    \\
			
			methods for solving the attention score & QKSAS    &  weak measurement & QLS  & Gaussian projection      \rule{0pt}{12pt} \\
			
			data types for attention scores & probability vector  & scalar & tensor & scalar    \rule{0pt}{12pt} \\

			
			\bottomrule
		\end{tabular}
\end{table*}

\subsection{Experimental Setting}

The specific experimental configurations for AmHE, AnHE, AmQAOA, AnQAOA, and QSAN \cite{0.901} are explicated in Tab. \ref{Experimental comparison}, listing crucial parameters for the ansatzes and classical optimizers. In terms of ansatzes, vital indicators include the number of parameters in the trainable layer $U$, the number of layers in ansatzes, and the requisite qubits. In classical optimization, factors such as learning rate, loss function type, batch size, maximum step size, optimizer type, and associated parameters are carefully taken into account. Furthermore, some settings, such as the type and configuration of the classical optimizer and the number of training parameters of the ansatzes, are as similar as possible to ensure and emphasize a fair evaluation of the performance among the quantum models. Further, a relatively large learning rate is uniformly chosen to expedite convergence.

\subsection{Experimental Analysis}

\subsubsection{Classification Experiment}
The classification results of MNIST and Fashion MNIST without circuit noise for the five models are displayed in Fig. \ref{MNIST} and Fig. \ref{lr}, where QSAN is a model closely related to our study for comparative analysis while the other four are subsets of QKSAN. 
Fig. \ref{MNIST} (or Fig. \ref{lr}) consists of three subfigures that share the horizontal axis which has a maximum range of 120. The vertical axis represents the three key indicators, including test accuracy, train accuracy, and cost function, respectively.  As the classical parts are same, the above three key parameter indicators  above are fundamentally subject to the disparities on quantum models. Moreover, AmHE, AnQAOA, AmQAOA, AnHE, and QSAN are marked by blue, green, red, orange, and purple lines, respectively. Finally, the following conclusions can be drawn.

\begin{itemize}
	\item In accuracy, by counting the averages of the last 10 steps and using 11 parameters, AmHE (blue line) achieves the highest test accuracy and training accuracy at 99\% and 99.06\% in MNIST, as well as 98.05\% and 98.52\% in Fashion MNIST. This effectively demonstrates that the above sub-model of QKSAN can return high accuracy with few training parameters, which is a potential quantum advantage over the parameter magnitude of classical machine learning models.
	\item Regarding convergence speed, AmHE trails QSAN by 42 steps in the MNIST task and by 50 steps in the Fashion MNIST experiment. Nevertheless, it boasts the smallest loss function value, indicating superior learning capability. This emphasizes that ansatzes essentially determine the speed of convergence, the classification effect, and the learning ability.
\end{itemize}

 Finally, to summarize the noise-free experimental results, even if in the same framework, the quantum circuit structure profoundly affects the training effect, learning ability, and sensitivity of parameters, but it is striking that the QML models are able to exchange a small number of parameters for high accuracy returns, which seems to be a potential learning advantage. In addition, AmHE, a subclass of QKSAN, exhibits a stronger learning capability with fewer qubits and circuit layers compared to QSAN, albeit with a slower convergence speed.

 \subsubsection{Quantum Kernel Self-Attention Score of Amplitude Encoding Hardware-Efficient Ansatz}

 From experiment (1), AmHE is the relative superior among the four models. Its QKSAS for MNIST and Fashion MNIST at the first and last iterations are plotted in Fig. \ref{QAttenscore}, demonstrating the transformation of  QKSAS before and after training. In Fig. \ref{QAttenscore}, the horizontal axis os QQS, the vertical axis is QKS, and  the scale of each coordinate is the quantum basis vector $\{|c\rangle \}_{c=0}^{3}$. 
 What is more, QKSAS exponentially expands with the number of qubits in the first quantum register, e.g., measuring 2 qubits in the Fig. \ref{ansatz} yields a $4 \times 4$ QKSAS. To sum up, the assertion is as follows.
 \begin{itemize}
 	\item  The size of QKSAS matrix exponentially expands with the number of qubits from the first quantum register, hence its representation space is much larger than the classical one to accommodate more information.
 \end{itemize}
 
\subsubsection{Noise Experiments}
AmHE is one of the more outstanding of the QKSAN subclasses and thus is once again chosen for investigation. The noise experiment are depicted in Fig. \ref{Noise_experiments} to explore the link between the learning ability and noise immunity of AmHE. 
Here, the same settings as described in Tab. \ref{Experimental comparison} is used, but with different initial parameters. The training of AmHE is then restarted to obtain the results in Fig. \ref{Noise_experiments}. 
\begin{itemize}
	\item The results demonstrate  that AmHE has less variation in learning accuracy when subjected to bit flip error and amplitude damping error with a probability of 0.1, but the learning accuracy experiences a considerable decline as the error probability increases. This reflects that AmHE has some robustness to noise. 
	\item Furthermore, the AmHE loss function variations display greater resilience to amplitude damping error, resulting in a less significant decline in learning ability compared to bit flip error.
\end{itemize}

\subsubsection{Method Comparison}
Finally, a concise theoretical comparison is drawn between QKSAN and the methods \cite{0.900,0.901,0.902} listed in Section \ref{QAM}, as depicted in Tab. \ref{Method Comparison}.

From an implementation standpoint, QKSAN and QSAN exhibit the advantage of deploying seamlessly on a quantum computer in a single step. This facilitates the concurrent generation of quantum self-attention scores and additional outputs such as labels, without subsequent processing. It also means that quantum computers take on more tasks and perform them more thoroughly.

Moreover, in Tab. \ref{Method Comparison}, disparities in the definition of quantum self-attention scores result in divergent data formats for these scores. Strikingly, QKSAN and QSAN extend the scalar representation of the self-attention score into the probability vector and the tensor. This broadening of data types not only aligns the self-attention score with the mathematics of quantum systems but also elevates its spatial representation ability.

The comparison highlights that QKSAN and QSAN share similar theoretical advantages. However, in Tab. \ref{Experimental comparison}, it becomes evident that QKSAN exhibits greater efficiency in quantum resource utilization. Specifically, QKSAN necessitates only two quantum registers, whereas QSAN requires four when employing the same encoding method.

\section{Conclusion}\label{sec5}

A QKSAM that integrates the advantages of SAM and QKM is proposed to equip current QML models with a larger data representation space and extract important information from quantum data. Subsequently, a QKSAN framework is built based on QKSAM, which frees up half of the quantum resources by DMP and conditional measurements, further boosting applicability and feasibility.  In the end, four sub-models of QKSAN are deployed on PennyLane and IBM Qiskit platforms for emphasizing the decisive impact of quantum models on classification accuracy, convergence speed, learning rate robustness, and learning capability.  During this period, a potential value that more than 98.05\% high accuracy can be traded for very few parameters which are far less in magnitude than those of classical machine learning models, is uncovered in QKSAN. In addition, in the noise experiment, the best-performing AmHE is robust to noise to a certain extent, but shows more resilience to amplitude damping errors. At last, a comparison of the methods leads to the conclusion that QKSAN is an efficient quantum method to compute the quantum self-attention score and output in one step, which is more quantum resource-efficient than its counterpart QSAN. Our approach contributes to the QML model to pay more attention to the important parts of quantum data, laying the foundation for future quantum computers to process massive amounts of high-dimensional data.

\ifCLASSOPTIONcompsoc


\begin{IEEEbiographynophoto}{Ren-Xin Zhao}(Member, IEEE) received his B.S. degree from the College of Automation, Hangzhou Dianzi University, Hangzhou, China in 2017 and his M.S. degree from the College of Electrical and Information Engineering, Hunan University, Changsha, China in 2020. He is now a PhD student in the  School of Computer Science and Engineering at Central South University, Changsha, China. His interests include quantum machine learning, quantum neural networks, and design and optimization of quantum circuits.\end{IEEEbiographynophoto}

\begin{IEEEbiography} [{\includegraphics[width=1in,height=1.25in,clip,keepaspectratio]{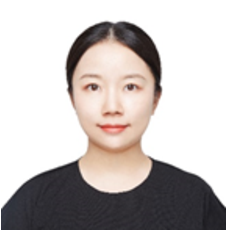}}]{Jinjing Shi}(Member, IEEE) is now an associate professor in the School of Electronic Information of Central South University. She received her B.S. and Ph.D. degrees in the School of Information Science and Engineering, Central South University, Changsha, China, in 2008 and 2013, respectively. She was selected in the ”Shenghua lieying” talent program of Central South University and Special Foundation for Distinguished Young Scientists of Changsha in 2013 and 2019, respectively. Her research interests include quantum computation and quantum cryptography. She has presided over the National Natural Science Foundation Project of China and that of Hunan Province. There are 50 academic papers published in important international academic journals and conferences. She has received the second prize of natural science and the outstanding doctoral dissertation of Hunan Province in 2015, and she has received the Best Paper Award in the international academic conference MSPT2011 and Outstanding Paper Award in IEEE ICACT2012.\end{IEEEbiography}

\begin{IEEEbiographynophoto}{Xuelong Li} (M'02-SM'07-F'12) is a full professor with School of Artificial Intelligence, OPtics and ElectroNics (iOPEN), Northwestern Polytechnical University, Xi'an 710072, P.R. China.
\end{IEEEbiographynophoto}


\ifCLASSOPTIONcaptionsoff
  \newpage
\fi

\end{document}